# The Parsec-Scale Radio Structure of NGC 1068 and the Nature of the Nuclear Radio Source

Jack F. Gallimore[1], Department of Physics, Bucknell University, Lewisburg, PA 17837, jgallimo@bucknell.edu
Stefi A. Baum and Christopher P. O'Dea, Space Telescope Science Institute, 3700 San Martin Dr., Baltimore, MD, sbaum@stsci.edu & odea@stsci.edu.

## ABSTRACT

We present sensitive, multifrequency Very Long Baseline Array (VLBA[2]) images of the nuclear radio sources of NGC 1068. At 5 GHz and 8.4 GHz, the radio continuum source S1, argued to mark the location of the hidden active nucleus, resolves into an elongated, ~ 0.8 pc source oriented nearly at right angles to the radio jet axis but more closely aligned to the distribution of the nuclear $H_2O$ maser spots. S1 is detected at 5 GHz but not at 1.4 GHz, indicating strong free-free absorption below 5 GHz, and it has a flat spectrum between 5 GHz and 8.4 GHz. A 5—8.4 GHz spectral index map reveals an unresolved, inverted spectrum source at the center of the S1 structure which may mark the AGN proper. The average brightness temperature is too low for synchrotron self-absorption to impact the integrated spectrum significantly. In addition, a careful registration with the nuclear $H_2O$ masers argues that the S1 continuum source arises from the inner regions of the maser disk rather than a radio jet. The emission mechanism may be direct, thermal free-free emission from an X-ray-heated corona or wind arising from the molecular disk. We demonstrate that the hidden active nucleus is sufficiently luminous, to within the current estimates, to provide the requisite heating. The radio jet components C and S2 both show evidence for free-free absorption of a compact, steep-spectrum source. The free-free absorption might arise from a shock cocoon enveloping the compact radio sources. The presence of $H_2O$ masers specifically at component C supports the interpretation for the presence of a jet-ISM interaction. Component NE remains a steep-spectrum source on VLBA baselines and appears to be a local enhancement of the synchrotron emissivity of the radio jet. The reason for the enhancement is not clear; the region surrounding component NE is virtually devoid of narrow line region filaments, and so there is no clear evidence for interaction with the surrounding ISM. Component NE might instead arise in an internal shock, or perhaps denser jet plasma that broke away from an earlier interaction with the circumnuclear ISM.
.

[1] Presently on leave at the Space Telescope Science Institute
[2] The VLBA is operated by the National Radio Astronomy Observatory. The National Radio Astronomy Observatory is a facility of the National Science Foundation operated under cooperative agreement by Associated Universities, Inc.

# 1 Introduction

The classical Seyfert 2 galaxy NGC 1068 has stood as the archetype for Seyfert unifying schemes ever since the discovery of strongly polarized optical continuum and broad line emission (Miller & Antonucci 1983; Antonucci & Miller 1985). X-ray spectra revealed Fe K emission with large equivalent width (Koyama et al. 1989; Marshall et al. 1993; Smith, Done, & Pounds 1993; Ueno et al. 1994), which is a clear indication that the active nucleus is hidden from direct view by a dense obscuring medium (Krolik & Kallman 1987). It appears that even hard X-rays are detected only in reflection (Matt et al. 1997), and the obscuring medium must therefore be Thomson thick.

There has been some discussion regarding the size and nature of the obscuring medium, whether, for example, molecular clouds distributed on scales of hundreds of parsecs (Cameron et al. 1993; Helfer & Blitz 1995; Schinnerer et al. 2000) or a compact, parsec-scale molecular torus which might ultimately feed an accretion disk (Krolik & Begelman 1986; Krolik & Begelman 1988). It has however been difficult to study the properties of the putative torus owing to the small angular scale; assuming a distance of 14.4 Mpc (Bland-Hawthorn et al. 1997), 1 parsec corresponds to 14 milliarcseconds (mas). Moreover, it has proven difficult to pin down the exact location of the central engine because it is so heavily obscured.

Studies of luminous $H_2O$ maser emission have provided the most detailed picture of the molecular ISM on sub-parsec scales. The brightest maser spots are associated with radio continuum component S1 (Gallimore et al. 1996b; Gallimore et al. 1996c; Gallimore et al. 2001) (see Figure 1 for orientation). The S1 masers resolve into a linear distribution of spots oriented nearly at right angles to the local radio jet axis (Greenhill & Gwinn 1997; Gallimore et al. 2001). The kinematics of the maser spots indicates that the masers are located in a rotating disk with inner radius ~ 0.65 pc and outer radius ~ 1.1 pc (Greenhill & Gwinn 1997). The mass enclosed within the inner radius is ~ $10^7$ $M_\odot$, marking the S1 maser source as the location of the active nucleus. It should be noted that the maser disk appears to be geometrically thin, however, and it is not clear how much it contributes to the nuclear obscuration.

There is additional evidence to support the identification of component S1 as the location of the central engine. NGC 1068 harbors a kiloparsec-scale, steep-spectrum radio jet (e.g., Wilson & Ulvestad 1987), but component S1 has an inverted spectrum between 5 GHz and 22 GHz (Gallimore et al. 1996c; Muxlow et al. 1996). Studies of the optical—UV polarization pattern and careful astrometric registration also point to S1 as the source of the scattered nuclear continuum (Capetti, Macchetto, & Lattanzi 1997b; Kishimoto 1999). S1 is coincident with the most powerful infrared source of NGC 1068 (Braatz et al. 1993; Bock et al. 2000). The source size is 2—3 pc in the near—mid infrared (Thompson & Corbin 1999; Jaffe et al. 2004), and the emission mechanism is almost certainly thermal radiation from hot (~ 1000 K) dust grains (Ward et al. 1987; Thatte et al. 1997). The presence of the compact, near—mid infrared source associated specifically with the AGN is arguably the best evidence for a compact obscuring medium in NGC 1068.

Observations with the Very Long Baseline Array (VLBA) resolve the radio continuum structure of S1 (Gallimore, Baum, & O'Dea 1997b; Roy et al. 1998, hereafter, RCWU), and the results are peculiar. Firstly, although it has a flat or inverted radio spectrum, the average brightness temperature is too low for synchrotron self-absorption to have an effect on the spectrum. Secondly, the radio source extends in a direction nearly at right angles to the radio jet axis but more closely aligned with the $H_2O$ maser disk. We had interpreted the radio source as arising from hot, photoionized gas on the inner, exposed edge of a parsec-scale obscuring torus. According to this interpretation, the emission mechanism could have been direct thermal free-free emission or electron-scattered non-thermal emission originating from the obscured central engine. RCWU also considered the possibility of synchrotron emission from the torus, although it was left unclear (1) how to generate synchrotron particles in the molecular medium and (2) how also to flatten the radio spectrum. Further complicating the interpretation, the VLBA maser observations of Greenhill & Gwinn (1997) were not phase-referenced, and so the astrometry of the $H_2O$ masers relative to the VLBA continuum emission was not well-constrained.

In an effort to understand better the nature of the nuclear radio source S1, as well as the other compact radio sources in the hundred parsec-scale nuclear jet, we present sensitive, new multifrequency VLBA and MERLIN observations of NGC 1068. We also employed updated data reduction techniques to improve the image quality of our earlier 8.4 GHz observations (from Gallimore et al. 1997b). Finally, we have performed a careful registration of the $H_2O$ maser spot positions relative to the VLBA continuum images of S1. These observations and analyses have provided better measurements of the radio continuum spectra of the nuclear radio components and further pin down the association between the masers and radio continuum associated with S1.

This paper is organized as follows. Section 2 provides an overview of the observations and data reduction. In section 3, we discuss the techniques used to measure and map the spectral indices of the compact radio continuum sources, the effects of spatial filtering on the spectral index measurements, and the careful registration of the $H_2O$ maser positions. Section 4 compares the present results with our earlier work and the measurements of RCWU. In Section 5, we consider the results for each of the compact radio sources in greater detail, but we pay particular attention to the nuclear radio source S1 and provide a more detailed analysis of the thermal free-free model for the radio continuum emission. Section 6 concludes with a brief summary of the main results of this work.

# 2 Observations and data reduction

The basic observing parameters of the new VLBA and MERLIN observations and image properties are summarized in Table 1. With MERLIN we observed 5 GHz continuum, and with the VLBA we observed continuum at three frequencies, 1.4 GHz (L-band), 5.0 GHz (C-band), and 8.4 GHz (X-band). Each VLBA observation was augmented by the VLA operating in phased-array mode to improve the calibration and imaging sensitivity. All observations were taken simultaneously in two polarizations (right-circular and left-circular polarizations; cross-polarizations where available were discarded), and all of the images generated from the calibrated data are total intensity (Stokes I) maps. Each of the observations required unique attention towards successful data reduction and imaging, however, and we therefore consider the particular details of the observations with each telescope array and each frequency band separately in the discussion that follows.

## *2.1 MERLIN 5 GHz observations*

We observed NGC 1068 using the 6-antenna MERLIN array on 2 January 1998. Observations of NGC 1068 were interleaved every 7 minutes with ~ 1 minute scans of the phase calibrator J0239−0234. The data were then calibrated and imaged using standard procedures within the AIPS software package (van Moorsel, Kemball, & Greisen 1996). The telescope gains were set by short-baseline observations of the flux standard 3C 286 which were bootstrapped to observations of the point calibrator OQ208 ($2.34 \pm 0.02$ Jy). The pointing center for scans of the phase calibrator J0239−0234, located 2°39′ from NGC 1068, were offset to the position reported by Muxlow et al. (1996). That position was known to about 20 mas accuracy, but the VLBA observations (discussed below) used a more accurate position, good to about 0.8 mas, from the ICRF catalog of Ma et al. (1997). We therefore calibrated the MERLIN data against the 5 GHz VLBA image of J0239−0234 obtained as part of the VLBA calibration (discussed below) to ensure that the MERLIN astrometry agreed as well as possible with the VLBA astrometry.

After gain and phase calibration, the source data were imaged by fourier inversion, CLEAN deconvolution, and self-calibration based on the resulting CLEAN component model. The MERLIN 5 GHz image is shown in Figure 1. Formally, the restoring beam size is $105 \times 51$ mas elongated along PA 20°, but, for the purpose of comparison with the earlier MERLIN image of Muxlow et al. (1996), we restored the CLEAN components with a 60 mas circular restoring beam. The structure of the radio continuum is simple enough that no significant artifacts should have been introduced by this slight super-resolution.

## 2.2 VLBA 1.4 GHz (L-band)

We observed NGC 1068 with the VLBA at 1.4 GHz (L-band) on 28 April 1997. We anticipated (and found) that NGC 1068 would be too faint, and its radio structure too complicated, initially to self-calibrate the interferometric phase, particularly on the longest baselines of the VLBA. We therefore employed a phase-referencing strategy using the calibrator source J0239–0234, the same calibrator used by Muxlow et al. (1996), RCWU, and the MERLIN 5 GHz measurements described above. The VLBA stations and phased VLA were nodded between the calibrator and target positions with integration times 6 minutes on target and 2 minutes on calibrator. The observations also included 8-minute scans on standard fringe finders primarily for phase alignment between IF channels.

The data were reduced using AIPS software and following standard procedures recommended by NRAO (Diamond 1995; Ulvestad 2001). The flux scale for the VLBA antennas was set based on telescope gain and system temperature monitoring data provided by NRAO. The flux scale of the phased-VLA was set by amplitude self-calibration based on images of the calibrators generated from data involving only the VLBA antennas. This bootstrapped amplitude calibration for the VLA was checked by producing new images of the calibrators including data from the VLBA and phased-VLA antennas. The resulting images showed no significant artifacts that would be expected for single-antenna amplitude errors, and the recovered flux on the calibrator images changed by less than 2% after including the phased VLA.

The visibility phases were initially calibrated for instrumental offsets between intermediate frequency channels based on fringe calibrator observations. We then applied fringe fitting to the phase reference calibrator to solve for phase and fringe rate corrections. The calibration solutions were interpolated in time and applied to the NGC 1068 visibility data.

We initially produced a naturally weighted image by Fourier inversion and standard CLEAN deconvolution (Högbom 1974; Schwarz 1978; Clark 1980; Schwab 1984). Although the components NE, C, and S2 were clearly detected, the initial image was plagued by sidelobe artifacts, in particular, negative "bowls" surrounding the sources that result from poor recovery of extended emission. We re-imaged the data using the multi-resolution CLEAN algorithm (Wakker & Schwarz 1988) that has recently been introduced to the AIPS task "IMAGR." Multi-resolution CLEAN improved the images by mitigating the sidelobe artifacts, and we used the multi-resolution clean components to self-calibrate the visibility phases; formally, the background noise fell slightly from 30 μJy $beam^{-1}$ to 20 μJy $beam^{-1}$ with the advent of the multi-resolution technique. The final, multi-resolution CLEAN image is provided in Figure 2.

### 2.3 VLBA 5.0 GHz (C-band)

The 5.0 GHz (C-band) VLBA observations took place on 26 April 1997. The observing and primary calibration strategies were identical to those employed for the L-band observations, except for self-calibration which involved CLEAN deconvolution and model fitting in the software package DIFMAP (Shepherd 1997). The main, compact radio components, NE, C, S1, and S2 were readily detected (even before self-calibration) and all are well-resolved. We found that, as was the case for the L-band data, the multi-resolution CLEAN outperformed the traditional implementation of CLEAN at recovering the extended structure. The final, multi-resolution CLEANed images of each component are provided in Figure 3.

### 2.4 VLBA 8.4 GHz (X-band)

The 8.4 GHz (X-band) observations took place on 6 May 1996, preceding the L-band and C-band observations by nearly a year. We originally self-calibrated the data using the standard technique involving CLEAN deconvolution and an antenna-based, least-squares correction of the interferometric phases; the resulting image of radio component S1 was published in Gallimore et al. (1997a; 1997b). We improved the recovery of extended emission in these data by employing the same DIFMAP self-calibration and multi-resolution CLEAN deconvolution described above for the C-band data. The image background noise was reduced by nearly a factor of 2 owing to the improved recovery of short spacing flux. To facilitate a comparison with the C-band data, we tapered the X-band data to the maximum baseline separation within the C-band data ($D / \lambda = 140 \times 10^6$) and, convolved the multi-resolution CLEAN components to the C-band CLEAN beam. The new, tapered images are given in Figure 4.

## 3 Analysis

### 3.1 Radio continuum spectra and short spacings issues

The VLBA is a fixed array and, as such, does not permit exact matching of the ($u$, $v$) coverage for data obtained at different radio frequencies. This issue is partly ameliorated by the robustness of image fidelity to the loss of intermediate spatial frequencies (e.g., Hecht 2001). Nonetheless, a main goal of the present work is to measure the radio continuum spectra of the individual components and map the spectral indices of resolved sources. Mismatched ($u$, $v$) coverage may impact these measurements, and this section presents our efforts to evaluate and correct any resulting biases.

To summarize the main properties of each radio continuum component, we calculated image moments using a modified version of the AIPS task 'MOMFT.' The results are given in Table 2. Note that all source sizes reported in this table and the tables that follow have been corrected for beam-smearing. The uncertainties include quadrature contributions from the rms image noise, a 5% uncertainty of the calibration of the VLBA fringe amplitudes, and a 6.8% uncertainty owing to deconvolution effects, based on the analysis of Wilson et al. (1998). The integrated spectra, including corrected measurements described below, are plotted in Figure 5.

There are two effects that preclude accurate spectral index measurements based on the integrated flux measurements. Firstly, for a fixed telescope array, the short-spacing coverage degrades with increasing observing frequency. Extended emission will resolve out at higher frequencies, resulting in an artificially steeper spectrum. The second effect is confusion with extended, steep spectrum emission. This effect is more difficult to assess, because extended structures that are present at low observing frequencies but are missing at the higher frequencies may be either (1) steep-spectrum or (2) resolved out owing to differences in $(u, v)$ coverage. One indicator of such confusion is a steep spectrum at low frequencies and flatter spectrum of the spectrum at higher frequencies, as appears to have happened with component S2; see Section 5.3, below.

Ultimately, the two effects described above result from extended emission contributing to a compact source structure. To help ensure that the integrated spectrum between 5 GHz and 8.4 GHz compares flux densities over the same region of the sky, we re-measured the properties of 5 GHz images after masking out regions that were not detected on the 8.4 GHz image. The image moments of the masked 5.0 GHz images and the 5.0 − 8.4 GHz spectral indices are provided in Table 3. (We use the definition for spectral index, $\alpha$, $S_\nu \propto \nu^\alpha$, so that steep spectra have negative spectral index and inverted spectra have positive spectral index.) The masked 5 GHz measurements will include some extended emission owing to beam-smearing into the masked region, and so the measured flux densities should still be viewed as upper (but tighter) limits for comparison with the 8.4 GHz measurements. As a result, the spectral indices are lower limits.

These considerations apply similarly to spectral indices measured from the 1.4 and 5 GHz images. For these frequencies, we generated new images after properly filtering each dataset to a common range of $(u, v)$ spacings. This filtering removed nearly half of the 5 GHz data resulting in a factor of 50% increase of the rms map noise, owing partly to reduced dynamic range. In contrast, only about 4% of the 1.4 GHz data were filtered by this operation. The image moments and spectral indices of the $(u, v)$-filtered images are provided in Table 3.

We also calculated spectral index maps for each of the compact radio sources using the AIPS task 'COMB.' The results are provided in Figure 6. Regions where the surface brightness falls below $3\sigma$ based on the map rms were blanked prior to image combination. We also generated formal uncertainty images based on the map rms, and the uncertainties are indicated by varying levels of grayscale on Figure 6. The effect of flux calibration and deconvolution uncertainties introduce an additional overall uncertainty of order 0.1 – 0.2 dex (Table 3) over the entire spectral index map.

As discussed above, the mismatched (*u*, *v*) coverage may systematically and artificially steepen these maps. To investigate the impact of this effect, we simulated VLBA observations of a flat-spectrum source used the AIPS task 'UVMOD'. The model was a gaussian surface brightness distribution scaled to 10 mJy total flux density and the geometry of S1 on the (real) 5 GHz image. Random, gaussian noise was added to the visibilities at a level appropriate to produce an rms image noise comparable to that observed, and the simulated data were imaged using the same deconvolution techniques as we had used on the actual data. The resulting, simulated spectral index map, which should have shown uniformly 0 dex, instead produced regions steepened by 0.1 – 0.2 dex, comparable to the systematic uncertainties owing to flux calibration and deconvolution. We conclude that spatial filtering resulting from the aperture synthesis technique may have steepened the spectral index maps by 0.1—0.2 dex, and the indices mapped in Figure 6 are therefore lower limits to the true value.

## 3.2 Astrometry of the radio continuum and $H_2O$ maser spots

Greenhill and Gwinn (1997) used the VLBA to map the distribution of 22 GHz $H_2O$ sources ("spots") around component S1. Those data were self-referenced relative to the brightest maser spot, the absolute position of which was not precisely known. To recover the absolute astrometry, we aligned their spot maps to the VLBA continuum images by bootstrapping their relative spot positions to phase-referenced VLA observations (Gallimore et al. 2001); the locations of the maser spots relative to the radio continuum emission from S1 is shown in Figure 7. The alignment involved generating a cross-correlation map of the VLBA maser spot positions and velocities relative to the VLA data, with each data point blurred by a three-dimensional gaussian with the FWHM scaled to the positional uncertainty in either sky coordinate and the channel-width along the velocity coordinate. The cross-correlation map contained only a single maximum (i.e., no additional, comparably strong local maxima) and therefore a unique alignment was determined. The statistical uncertainty of the alignment, estimated by the FWHM of the cross-correlation peak, is ~ 0.9 × 0.5 mas, elongated along the major axis of the maser spot distribution. For comparison, the characteristic error of the VLA maser centroid positions is around 0.5 mas for the brightest masers. Because many VLBA maser spots fill the coarser VLA beam (80 mas) and channel-width (~ 10 km $s^{-1}$) for a given VLA channel, and the result is a smearing of the positional uncertainty along the major axis of the spot distribution. The radial velocity gradient determines the scale of the smearing: a single 10 km $s^{-1}$ VLA channel includes emission from a ~ 0.3 – 1 mas distribution of maser spots (Greenhill & Gwinn 1997), which agrees well with the uncertainty along the maser spot axis measured from the cross-correlation map.

Self-calibration introduced negligible uncertainty to the VLBA continuum images. After the first round of self-calibration, the centroid of S1 on the 5 GHz image drifted only ~ 0.1 mas south and ~0.1 mas east, comparable to the uncertainty of the centroid measurement. Any astrometric drifts after three subsequent iterations were immeasurably small (i.e., << 0.1 mas). The astrometry of the 8.4 GHz image is consistent with the 5 GHz image and must therefore have comparably small astrometric errors owing to self-calibration.

There remains an additional uncertainty of the continuum astrometry owing to the accuracy of the VLBA antenna positions and atmospheric variations that are not accounted for in the phase calibration procedure for the VLBA data specifically. Wrobel et al. (2000) indicate that standard phase referencing procedures, such as were used here, introduce ~ 1 mas uncertainty into the astrometry (see also Lestrade 1991 and Beasley and Conway 1995). Accounting for this additional uncertainty, the overall uncertainty of the VLBA $H_2O$ maser positions relative to the VLBA continuum images is roughly 1.3 mas along the maser axis and 1.1 mas perpendicular to the maser axis.

# 4 Results: comparison with earlier VLBA measurements

All of the components show well resolved structure on these new VLBA images. The new images show broad agreement with the VLBA snapshots presented in RCWU, including the east-west orientation of component NE, which is also apparent on MERLIN and VLA images (Gallimore et al. 1996c; Muxlow et al. 1996), and the detection of a faint, southeastern extension of component C at 5 GHz (comparing to Figure 1 in RCWU). The longer integration time of the present observations, the improved ($u$, $v$) coverage that results, and inclusion of the phased-VLA has improved the sensitivity by a factor of 12 at both 1.4 GHz (compared to the 1.7 GHz data of RCWU) and 5 GHz. The improved sensitivity is particularly vital to the study of radio source S1 because of the possible thermal nature of the radio emission. In addition, the extra attention brought to optimal self-calibration and deconvolution techniques has also greatly improved the quality of the 8.4 GHz image of S1 (Figure 7) compared to our earlier effort (Gallimore et al. 1997b).

There are two immediate results owing to the improved sensitivity of these observations. The VLBA observations newly detect component S2 at 5 GHz with peak flux density just below the detection threshold reported by RCWU. We also recovered diffuse emission associated with component C at 1.4 GHz. Although the integrated flux density is 42 mJy (Table 2), the peak surface brightness is only ~ 2 mJy, again below the detection threshold of RCWU. There is no compact source that can be unambiguously associated with the brighter compact source seen in the 5 and 8.4 GHz images of component C, however. Consistent with the RCWU observation, that compact source must be severely attenuated at 1.4 GHz (see also Figure 5).

The new 5 GHz VLBA observations also recover more flux density from components NE and S1 than the observations of RCWU. Again the reason is almost certainly the improved sensitivity to low surface brightness emission. For example, the recovered flux density of component NE is 37 mJy, compared to 15 mJy in RCWU. The extra flux density arises from low surface brightness emission along the minor axis of component NE: the minor axis FWHM reported by RCWU is 10 mas, but we measure 17 mas (Table 2). Similarly, the new 5 GHz VLBA observations recover a total of 9.1 mJy from component S1 compared to 5.1 mJy from RCWU. The compact emission accounts for 5.9 mJy (Table 3), in reasonable agreement with RCWU. For comparison, the 5 GHz flux density of S1 based on a gaussian fit to the MERLIN image (Figure 1) is 12 ± 1 mJy (see also Muxlow et al. 1996).

There remains one discrepancy between our measurements and RCWU that cannot be accounted for by sensitivity considerations: component C appears significantly fainter in our observations. RCWU report a 5 GHz flux density of 27 mJy, but we measure only 19 mJy. Since our VLBA observations recover significantly more flux for the other components, the discrepancy cannot be ascribed to an error of the flux scale calibration; rather, the more sensitive observations should have recovered at least as much flux density. The simplest conclusion is that the mas-scale structure of component C is variable, perhaps owing to the evolution of the shock structure as the jet interacts with the molecular cloud that produces the local $H_2O$ maser emission (see the discussion in section 5.2, below). The time between measurements is 331 days, implying a source diameter ≤ 0.6 pc, corresponding to an angular diameter ≤ 8 mas. Allowing for the uncertain distance to the source, this limit is compatible with the compact source size measured on the 8.4 GHz image, 8.8 × 3.1 mas (Table 2).

# 5 Discussion

## 5.1 Component S1

S1 almost certainly marks the location of the nucleus. The integrated radio spectrum is inverted below 5 GHz and flat above 5 GHz (Gallimore, Baum, & O'Dea 1996a; Gallimore et al. 1996c; Muxlow et al. 1996; Gallimore et al. 1997b) (see also Figure 5 and Table 3). The brightest $H_2O$ maser emission associates with S1 (Claussen & Lo 1986; Gallimore et al. 1996b; Gallimore et al. 2001), and we confirm that the masers are closely associated with the compact, mas-scale structure (Figure 7). The kinematics and geometry of the masers are compatible with a parsec scale, thin rotating disk surrounding a $1.2 \times 10^7$ $M_\odot$ compact mass, presumably the AGN (Greenhill et al. 1996; Greenhill & Gwinn 1997; Gallimore et al. 2001). Polarization studies of the scattered AGN light coupled with careful registration of the radio and optical coordinate reference frames also points to S1 being coincident (within ~ 0.1” uncertainty) with the scattering center (Capetti et al. 1997b; Kishimoto 1999).

The non-detection at 1.4 and 1.7 GHz indicates strong attenuation below 5 GHz. At 5 GHz, $T_b \sim 2.5 \times 10^6$ K, too low for synchrotron self absorption to affect the spectrum, unless the magnetic fields are in the $10^9$ gauss regime (Gallimore et al. 1996a). RCWU considered Razin-Tsytovitch suppression and ionization losses of a synchrotron source mixing with thermal plasma. Assuming equipartition conditions in the putative synchrotron plasma, the critical frequency for Razin suppression is 22 MHz, and ionization losses furthermore cannot sufficiently steepen the spectrum below 5 GHz. Free-free absorption is the remaining and likeliest explanation. At GHz frequencies, the free-free opacity of a foreground, isothermal plasma can be approximated by

$$\tau_\nu^{ff} \approx 0.08235 \times T_e^{-1.35} \nu_{GHz}^{-2.1} \left\langle n_e^2 \right\rangle L_{pc}, \tag{1}$$

where $T_e$ is the electron temperature, $\nu_{GHz}$ is the radio frequency in GHz, $<n_e^2>$ is the mean square electron density in $cm^{-6}$, and $L_{pc}$ is the path length along the sight-line in parsecs (Altenhoff et al. 1960; Mezger & Henderson 1967). Assuming $T_e \sim 10^6$ K (see the discussion below), an emission measure EM = $\langle n_e^2 \rangle L > 3 \times 10^9$ $cm^{-6}$ pc suffices to attenuate 1.4 GHz emission. Models of X-ray irradiated molecular tori predict such high emission measures and free-free opacity (Krolik & Lepp 1989; Neufeld, Maloney, & Conger 1994), and so the presence of $H_2O$ masers and high free-free opacity appears to be consistent with S1 marking the location of the AGN.

### 5.1.1 Geometry and Emission Mechanisms

The parsec-scale, resolved continuum emission detected at both 5 and 8.4 GHz agrees both in position and reasonably in orientation with the $H_2O$ maser spots. The east-west extent of the continuum source is somewhat more compact than the maser spot distribution; based on the position-velocity profile, the inner radius of the maser disk is ~ 0.6 pc, but the semi-major axis of the 8 GHz continuum image is ~ 0.4 pc. The geometry suggests that the radio continuum arises from the inner part of the molecular (maser) disk rather than the radio jet (Gallimore et al. 1997b). Radio source detections on VLBA-scale baselines are usually associated with synchrotron emission, but it is not clear how to generate synchrotron emission in a parsec-scale disk (see, e.g., the discussion of RCWU).

We prefer the interpretation that the radio source traces direct, thermal free-free emission from the X-ray irradiated molecular disk (Gallimore et al. 1996a, 1997a). Figure 8 schematically illustrates the geometry that we propose. The main advantages of this interpretation are (1) free-free emission can produce the flat or inverted spectrum observed over the GHz frequency range (Gallimore et al. 1996a), and (2), unlike synchrotron emission, free-free emission would naturally arise from X-ray-heated disk gas. This model may also explain the spectral index structure; if the volume filling fraction is near unity (i.e., the geometry is more like a disk and less like an annulus), the optical depth through the center of the disk is greater than through the edges. The spectral index should therefore be more inverted toward the center and flatter away from center, as is observed (Figure 6).

Assuming that free-free emission dominates the centimeter-wave spectrum of S1, the basic properties of the thermal plasma can be estimated from the observed brightness temperature and source size. Averaged over the inner (masked) region of the 5 GHz image, $\langle T_b \rangle = 4 \times 10^6$ K. The spectrum delineated by the three frequency bands is consistent with the plasma becoming optically thin above 5 GHz. Adopting $\tau_{ff} = 1$ at 5 GHz and a slab geometry, the characteristic electron temperature is $T_e = 6 \times 10^6$ K, which is near the estimated Compton equilibrium temperature (Krolik & Begelman 1986; Kinkhabwala et al. 2002), accepting uncertainties in the shape of the ionizing spectrum. The source major axis is ~ 0.8 pc, which we adopt as a characteristic path-length through the plasma. Applying the free-free opacity approximation appropriate for GHz frequencies, Equation (1), the rms electron density is $n_e \sim 8 \times 10^5$ cm$^{-3}$. For comparison, the Thomson opacity for these estimates of the plasma conditions is $\tau_e \sim 1.3$, slightly in excess of the free-free opacity. Thomson scattering effectively increases the free-free opacity by increasing the path length required for radiation to escape. Accounting for this opacity enhancement slightly reduces the requisite electron density to $n_e \sim 6 \times 10^5$ cm$^{-3}$. Assuming a cylindrical symmetry and uniform density, the mass of the ionized gas is inferred to be ~ 3400 $M_\odot$.

### 5.1.2 The luminosity of the hidden AGN

The main concern with this interpretation is whether the hidden AGN is sufficiently luminous to heat dense, disk plasma to near the Compton equilibrium temperature (i.e., is the ionization parameter sufficiently high at a radius of a few tenths of a parsec?). Unfortunately, the central engine is hidden by foreground dust, and so the actual luminosity cannot be measured directly. The Eddington luminosity, or $L_{edd}$, sets the upper limit. The maser disk kinematics argue for the presence of a ~ $10^7$ $M_\odot$ black hole (Kumar 1999; Huré 2002; Lodato & Bertin 2003), and so $L_{edd} \sim 10^{45}$ ergs s$^{-1}$.

The shape of the intrinsic spectral energy distribution (SED) of the AGN is detected only in reflection from surrounding clouds (Pier et al. 1994). Luminosity measurements based on the reflected spectrum therefore depend on $f_{refl}$, the fraction of the SED that is reflected into our sight-line. Estimates of $f_{refl}$ range over an order of magnitude, $0.1\% \leq f_{refl} \leq 1.5\%$ (Baldwin, Wilson, & Whittle 1987; Bergeron, Petitjean, & Durret 1989; Bland-Hawthorn, Sokolowski, & Cecil 1991; Miller, Goodrich, & Mathews 1991; Storchi-Bergmann, Mulchaey, & Wilson 1992; Bland-Hawthorn & Voit 1993; Pier & Krolik 1993; Iwasawa, Fabian, & Matt 1997). Based on the analysis of Pier et al. (1994), this range of $f_{refl}$ constrains the bolometric AGN luminosity to $2.4 \times 10^{44}$ ergs s$^{-1}$ $\leq L_{AGN} \leq 3.6 \times 10^{45}$ ergs s$^{-1}$. For comparison, Iwasawa et al. (1997) argue for an X-ray luminosity of order $10^{44}$ ergs s$^{-1}$.

Infrared radiation from the dusty, obscuring medium offers an independent measure of $L_{AGN}$. Bock et al. (2000) (and references therein) measure the infrared luminosity of the central 4″ to be $1.1 \times 10^{45}$ ergs s$^{-1}$ between 2 μm and 158 μm. They argue however that the radiation is highly beamed along the jet axis, and clouds in that direction may see an effectively higher luminosity (cf. Baldwin et al. 1987). The mid-infrared luminosity of the compact, 0.1″-scale core specifically between 7.7 μm and 18.5 μm is ~ $3 \times 10^{44}$ ergs s$^{-1}$. The reflected spectrum shows no significant reddening (Miller et al. 1991; Pier et al. 1994), however, and ionizing photons must escape to produce the narrow line region. The dust covering fraction must therefore be less than unity, and the infrared luminosities are lower limits to $L_{AGN}$.

Based on these infrared – X-ray measurements and the additional constraint of the Eddington luminosity, the luminosity of the central engine falls in the range $10^{44}$ – $10^{45}$ ergs $s^{-1}$, with infrared observations favoring the high end. In the face of these constraints, we next consider the AGN luminosity needed to heat the S1 plasma and produce the flux densities observed on VLBI-scales.

### 5.1.3 X-ray Heating Models for Component S1

We employed the photoionization code XSTAR (Kallman 2001) to calculate a grid of plasma models as a function of incident AGN luminosity and plasma density with the goal of finding the minimum AGN luminosity needed to produce the observed 8.4 GHz flux density, $S_{8.4}$ = 5.4 mJy. A model for the 8.4 GHz measurement was chosen over the 5 GHz measurement because the 8.4 GHz image also suffers less confusion with extended emission and so is arguably a better measurement of the continuum emission associated directly with the disk plasma.

The Pier et al. (1994) spectrum was used to describe the shape of the nuclear SED, which was extrapolated from 10 keV out to 100 keV assuming a photon index $\Gamma = -2$ (cf. Alexander et al. 2000), consistent with the BeppoSAX hard X-ray detection (Matt et al. 1997). The SED is shown in Figure 9. The disk plasma was modeled as having uniform density and an outer covering fraction of 34%, consistent with the imaging results. The outer radius was set to 0.4 pc also to match the 8.4 GHz image, and the inner radius was set to 0.01 pc to provide an order of magnitude radial scale to approximate a uniformly filled disk while avoiding computational singularities. XSTAR does not perform ionization and thermal balance calculations for photon energies below 0.1 eV, and so the radio flux densities are not directly predicted by the code. To recover the radio flux densities, we integrated the radiative transfer equation for the 8.4 GHz brightness temperature based on the radial profiles of temperature and electron fraction given by the XSTAR model and employing the free-free opacity approximation given in Equation (1). We then determined the integrated flux density assuming that the brightness temperature is uniform over the measured angular size at 8.4 GHz. This latter step is equivalent to replacing the edge-on disk geometry with a slab geometry, which is a good approximation in the optically thick limit. In support of this approximation, only models with $\tau_{ff}$(8.4 GHz) ≥ 1 were successful at reproducing the observed flux density.

Figure 10 provides the resulting grid of predicted 8.4 GHz flux density as a function of plasma density and incident AGN luminosity (integrated between 0.1 eV and 100 keV). The model behavior is consistent with that expected: greater radio flux density with increasing AGN luminosity (which results in more plasma at the Compton equilibrium temperature) and increasing plasma density (i.e., increasing free-free optical depth). From inspection of the model grid, the thermal interpretation for S1 requires $L_{AGN} \geq 7 \times 10^{44}$ ergs $s^{-1}$ to produce the observed 8.4 GHz flux density. This luminosity is a factor of only 1.8 greater than the Pier et al. (1994) estimate and therefore requires $f_{refl}$ = 0.6%, which lies within the observational constraints. The plasma density required at the minimum AGN luminosity is $5 \times 10^5$ cm$^{-3}$, comparable to the rough estimate given above. In addition, the predicted 1.4 GHz flux density is 0.4 mJy, consistent with the 0.7 mJy upper limit (Table 3).

Models that successfully produced the observed 8.4 GHz flux density of S1 also produced plasma heated to Compton equilibrium ($T_C = 3 \times 10^7$ K) out to $r \geq 0.3$ pc. The Compton temperature exceeds the escape temperature outside $r \approx 0.1$ pc, again assuming the presence of a $10^7$ $M_\odot$ black hole, and so the hot plasma would likely expand and cool adiabatically (Begelman, McKee, & Shields 1983; Krolik & Begelman 1986). This additional cooling would lower the temperature to a few × $10^6$ K (Begelman et al. 1983; Krolik & Begelman 1986), which would in turn increase the free-free opacity at GHz frequencies by a factor of ~ 10. These two competing effects of reduced electron temperature and increased free-free opacity roughly cancel to produce the same integrated brightness temperature. A more precise treatment would require the additional hydrodynamical modeling, which is beyond the scope of the present work.

The present analysis furthermore considers only photon heating (i.e., photoionization and Compton heating), but the AGN might provide other sources of energy, such as cosmic ray heating, that would reduce the AGN radiative luminosity requirements (Krolik, McKee, & Tarter 1981; Mathews & Ferland 1987). In addition, the disk would enjoy a relatively unobscured view of any reflected X-ray emission, which we did not include in the present models. The additional reflected component would contribute to disk heating out to larger radii (Haardt, Maraschi, & Ghisellini 1997; Nayakshin 2000; Nayakshin, Kazanas, & Kallman 2000) and again reduce the luminosity requirements. We conclude that a hidden AGN of luminosity $L_{AGN} \sim 7 \times 10^{44}$ ergs s$^{-1}$ could plausibly heat dense ($n_e \sim 5 \times 10^5$ cm$^{-3}$) gas to temperatures of order $10^6 - 10^7$ K within the central 0.4 pc of NGC 1068. Under these conditions, and assuming a covering fraction of ~ 34% (compatible with the VLBA images), the hot, dense gas could produce sufficient free-free continuum to be detectable on VLBA baselines.

### 5.1.4 A Comparison with the Soft X-ray Spectrum

The thermal free-free model for S1 predicts a rich soft X-ray line spectrum as illustrated in Figure 9. The S1 source lies within the obscuring dust that blocks our direct view of the nucleus, however, and so the optical—X-ray spectrum of S1 cannot be measured directly. On the other hand, the soft X-ray lines predicted by the model might be detectable in scattered light. The observed soft X-ray spectrum therefore places an additional constraint on the thermal model for S1: the predicted scattered spectrum should not exceed that observed.

Recent Chandra (Brinkman et al. 2002; Ogle et al. 2003) and XMM-Newton measurements (Kinkhabwala et al. 2002) failed however to detect any electron-scattered soft X-ray continuum emission from the AGN, perhaps owing to photoelectric absorption within the scattering medium. To quantify better the limits on the scattered light fraction in the soft X-ray band, we added and scaled the total (AGN continuum plus model S1 disk) spectrum that would be seen by the scattering mirror to the mean continuum level of the XMM-Newton RGS spectrum of Kinkhabwala et al. (2002). The results are shown in Figure 11. We chose to compare with the XMM-Newton data rather than Chandra because the larger aperture (15”) of the XMM-RGS measurement ensures that the entire electron scattering region is included, but the conclusions are not affected because the line equivalent widths do not change significantly in the smaller Chandra apertures. Note that radiative recombination continua from an extended hot component within the NLR is the primary source of the soft X-ray continuum, rather than electron-scattered AGN continuum (Young, Wilson, & Shopbell 2001; Brinkman et al. 2002; Kinkhabwala et al. 2002), and so the scaling factor that we calculated to match the XMM-Newton spectrum, 0.12%, is an upper limit on the soft X-ray reflection fraction.

The most luminous line in the X-ray irradiated disk model spectrum is O VIII Ly$\alpha$ ($\lambda$ 18.97 Å) (cf. Ballantyne, Ross, & Fabian 2002), and, after scattering, it appears to be weak but possibly detectable by current X-ray observatories (Figure 11). However, O VIII Ly$\alpha$ can contribute at most 30% of the observed line flux (set by the upper limit on the scattered soft X-ray continuum), and therefore the source of this line is probably a hot component of the NLR and not the disk. We conclude that the thermal free-free model for S1 is not ruled out by the soft X-ray observations; that is, the thermal free-free model does not overproduce soft X-ray line emission compared to the observations. Unfortunately, we also cannot confirm the thermal free-free model based on the predicted soft X-ray line emission owing to confusion with the hot NLR component.

### 5.1.5 Little contribution from a scattered, hidden radio source

As discussed above, the column densities required by the thermal free-free model predict a high electron scattering opacity. It therefore seems plausible that some fraction of the measured VLBA continuum might arise from a compact (i.e., unresolved), flat-spectrum synchrotron source that is hidden and blurred by the thermal electron scattering disk (Gallimore et al. 1996a). This model is attractive insofar as it may reduce the required contribution of thermal free-free continuum to the VLBA flux densities and thereby reduce the required AGN luminosity.

A detailed radiative transfer model for disk-scattered radio emission is beyond the scope of the present work, but we can estimate the conditions necessary for the scattering model by exploring the relevant opacities. Firstly, the surface brightness distribution of S1 is smooth with only a modest central concentration (Figure 4). Suppressing the contrast between the central source and the scattering disk requires a high scattering opacity, and so solutions only with $\tau_{es} > 1$ are valid for this model. Secondly, the free-free opacity should be less than unity, or else the scattered contribution would be squelched by free-free absorption, and the resulting continuum would be dominated by thermal free-free emission.

We computed grids of integrated $\tau_{ff}$(8.4 GHz) and electron scattering opacity, $\tau_{es}$, based on the XSTAR model runs; the results are provided in Figure 12. As expected, the scattering model favors both high $L_{AGN}$ and high $n_H$ to produce the high temperature and electron column necessary for $\tau_{es} > \tau_{ff}$. The scattering model requires $L_{AGN} \geq 5 \times 10^{44}$ ergs $s^{-1}$, which is not significantly reduced in comparison with the thermal free-free model as detailed above. In fact, based on comparison with Figure 10, scattering opacity dominates over the region of $L_{AGN} - n_H$ space where the flux density of thermal free-free emission exceeds the measured flux density of component S1. We conclude therefore that the conditions required for substantial scattering opacity produce excessive free-free continuum, and we can therefore rule this scattering model out. Any compact radio source hidden within the S1 disk must be suppressed by free-free absorption or must otherwise be weaker than the contribution of free-free emission at 8.4 GHz.

### 5.1.6 The residual spectrum

The MERLIN 5 GHz flux density of component S1 is 12 ± 1 mJy (Section 4). The VLBA recovers about 9 mJy at 5 GHz, of which about 6 mJy arises from the compact region associated with the $H_2O$ masers (Table 3). The 5 GHz flux density of the residual (MERLIN − VLBA disk), extended emission is 6 ± 1 mJy.

The VLA 22 GHz flux density of S1 is 19.1 ± 0.6 mJy (Gallimore et al. 1996c). Extrapolating the spectrum of the compact (disk) region based on our 5—8.4 GHz spectral index measurement, the compact region contributes 4.7 ± 1.8 mJy to the 22 GHz flux density. The predicted, residual (VLA − extrapolated VLBA) flux density of the extended region at 22 GHz is therefore ~ 15 ± 2 mJy. The size of the residual VLA source must be greater than 20 mas since it does not arise from the maser disk region (that component has been subtracted out by extrapolation), but it must be smaller than 80 mas since it is not resolved by the VLA.

Putting these measurements together, we predict that the 5—22 GHz spectral index of S1, after subtracting the contribution of the compact VLBA source, is $\alpha \sim +0.6$. The constraints on the source size limit the characteristic brightness temperature of the residual emission to $5.5 \times 10^3$ K $\leq T_b$ (22 GHz) $\leq 8.8 \times 10^4$ K. Like the compact source, the extended (> 20 mas) emission from S1 has an inverted spectrum and a brightness temperature that is too low for synchrotron self-absorption. Following the arguments above, free-free emission could plausibly generate this residual emission. The only requirements are that the electron temperature must be of order $10^4$ K, characteristic of AGN emission line regions, and the rms electron density must be at least $2 \times 10^4$ $cm^{-3}$ to ensure sufficient free-free opacity at 22 GHz.

Unfortunately, present aperture synthesis arrays cannot resolve the 22 GHz continuum emission with sufficient brightness temperature sensitivity to distinguish the compact and extended emission. Observations with the future New Mexico Array, a planned upgrade to the present VLA, may be able to test this prediction.

### 5.1.7 Interpretation: an X-ray heated wind

The present observations and analysis support the view that thermal free-free emission dominates the radio spectrum of component S1, but we cannot specifically exclude the possibility of non-thermal contributions to the 5 – 8 GHz flux density. Irrespective of the dominant emission mechanism, thermal or non-thermal, the morphology of the S1 VLBI source indicates the presence of hot plasma associated with the geometrically thin maser disk rather than a geometrically thick torus as we had earlier suggested (Gallimore et al. 1997b) or the radio jet. We speculate that a X-ray-heated wind arising from the inner part of the maser disk is the origin of the S1 radio emission. X-ray heating will produce a strong wind under the condition $T_{esc} / T_C > 1$, where $T_{esc}$ is the escape temperature and $T_C$ is the Compton temperature (Begelman et al. 1983; Ostriker, McKee, & Klein 1991). For NGC 1068, the conditions for a strong wind occur outside $r \sim 0.1$ pc, or one-fourth of the observed semi-major axis of the S1 radio source.

Characteristic wind speeds are expected to be of order a few hundred km $s^{-1}$ (Begelman et al. 1983; Pier & Voit 1995), and the mass loss rate is predicted to be of order 1 $M_{\odot}$ $yr^{-1}$ (Krolik & Begelman 1986). An upper limit on these rates can be set by assuming that the plasma expands into vacuum (i.e., in the limit where the disk plasma pressure greatly exceeds that of the environment). Under this condition, the initial expansion velocity is

$$v_{\exp} = c_s \left( \frac{2}{\gamma + 1} \right)^{1/2}, \tag{2}$$

where $c_s$ is the initial sound speed (Reynolds 1876; Lang 1999). At the Compton equilibrium temperature, the vacuum expansion velocity is $v_{exp}$ = 560 km $s^{-1}$. The corresponding mass loss rate is $\dot{M} \approx \rho v_{\exp} A$, where $A$ is the area of the face of the disk. Adopting $n_H = 5 \times 10^5$ $cm^{-3}$ based on the minimum $L_{AGN}$ model, $\dot{M} \approx 2$ $M_{\odot}$ $yr^{-1}$. As pointed out by Krolik & Begelman (1986) for their torus wind model, the fueling requirement for such an X-ray-heated disk wind greatly exceeds that required by the accretion disk to produce the X-ray luminosity; otherwise the disk evaporates on a time-scale of only $M / \dot{M} \sim 2000$ yr.

As the wind expands away from the central engine, the emission measure drops, and the gas cools adiabatically. As a result, the brightness temperature of the gas falls rapidly below the VLBA surface brightness sensitivity. Consistent with this picture, the 5 GHz VLBA image (Figure 3) shows fading extensions to the northeast and southwest along the symmetry axis of the maser disk. Thermal emission from cooler, expanded wind material is detectable by the VLA, and analysis of the residual 22 GHz emission is consistent with thermal emission from ~ $10^{4-5}$ K gas on scales of a few parsecs from the VLBA source.

It should be noted that the association of a VLBI continuum source with a disk rather than a jet is, so far, unique to NGC 1068 (e.g., Mundell et al. 2000). On the other hand, few relatively close and luminous Seyfert nuclei have been observed with similar sensitivity on VLBI baselines. NGC 1068 also stands out as one of the few extragalactic $H_2O$ megamaser sources, and, among those, one of the few galaxies harboring a compact $H_2O$ maser disk. It is perhaps not surprising that component S1 may be unique among Seyfert radio sources.

### *5.2 Component C*

The presence of $H_2O$ masers (Gallimore et al. 1996b; Gallimore et al. 2001) and a sudden bend of the radio jet (Gallimore et al. 1996c; Muxlow et al. 1996) provide evidence that component C marks a shock front in the collision between the jet and a molecular gas (Gallimore et al. 1996a; Gallimore et al. 1996b). Component C does not line up with any of the brighter NLR clouds (Capetti et al. 1997b; Kishimoto 1999; Thompson et al. 2001), but HST and Keck imaging has detected Pa$\alpha$ emission and mid-infrared dust continuum in close association (Bock et al. 2000; Thompson et al. 2001), indicating selective obscuration towards component C.

RCWU first noted the inverted spectrum of the VLBI-scale structure below 5 GHz and modeled the emission mechanism as intrinsically steep spectrum radiation (synchrotron emission) attenuated by free-free absorption through an “ordinary” cloud in the NLR. Their model was motivated in part by their alignment of component C with the bright NLR cloud C (perhaps based on Gallimore et al. 1996a). However, revised polarization studies now place component C roughly 0.3″ north of NLR cloud C into a region of weaker optical and infrared line emission (Capetti et al. 1997b; Kishimoto 1999). The RCWU model for the plasma conditions of the NLR clouds predicts an H$\beta$ luminosity comparable to or in excess of the integrated H$\beta$ luminosity of the brightest NLR clouds, and, in view of the revised astrometry, it seems that the NLR cloud model predicts too much H$\beta$ emission for the amount of free-free absorption required. Foreground dust extinction mitigates this problem, and the excess H$\beta$ may be reprocessed as the detected infrared continuum. Nevertheless, the NLR cloud model requires a chance alignment between an unseen NLR cloud and only the mas-scale structure of component C. In fact, there is no evidence for free-free absorption at GHz frequencies anywhere along the jet except specifically at the compact radio sources C, S1, and S2 (Gallimore et al. 1996c). Rather than requiring chance, mas-scale alignments uniquely between compact NLR clouds and the compact radio sources, it seems more natural to propose that the radio sources and free-free absorbing plasma share a common origin.

As occurs in Herbig-Haro objects (e.g., Bally et al. 1997), fast jets will dissociate and ionize molecular gas, whether by collisional processes or the absorption of ionizing radiation produced at the shock front (Dopita & Sutherland 1996), and so the shock may generate its own ionized envelope. To explain the low-frequency turnover of the radio spectra of compact radio sources (such as Gigahertz peak spectrum sources), Bicknell, Dopita, & O’Dea (1997) demonstrated that jet-induced shocks might naturally produce localized free-free absorption (see also O'Dea 1998). Bicknell et al. (1997) find that a 1000 km $s^{-1}$ shock velocity and a pre-shock density of 100 $cm^{-3}$ suffice to produce substantial free-free absorption at GHz frequencies. The advantage of this shock model over the NLR cloud model of RCWU is that it removes the need for coincidence; the jet-cloud shock that has produced component C could naturally produce a cocoon of free-free absorbing thermal plasma.

Based on our measurements (Table 3) and assuming an intrinsic (unabsorbed) spectral index $\alpha = -0.7$ below 5 GHz, the free-free opacity required to explain the diminution of the 1.4 GHz flux density is $\tau_{ff}(1.4\ \mathrm{GHz}) > 2.4$. Further assuming an electron temperature $T_4 = T_e / 10^4$ K, the emission measure through the absorbing plasma would have to be EM $= \langle n_e^2 \rangle L > 1.5 \times 10^7\ T_4^{1.35}\ \mathrm{cm}^{-6}$ pc to produce sufficient free-free absorption. For comparison, Dopita & Sutherland (1996) have calculated the conditions of jet-shocked NLR clouds over a range of parameter space appropriate for Seyfert galaxies. Their models predict electron columns through the either the post-shock plasma or shock precursor region $N_e = \langle n_e \rangle L \sim 10^{18} - 10^{22}\ \mathrm{cm}^{-2}$. Pre-shock molecular cloud densities provide an order-of-magnitude estimate of the post-shock or precursor electron density. Molecular clouds located within ~ 1″ of the nucleus have measured densities $n_{H2} \sim 10^6\ \mathrm{cm}^{-3}$ (Tacconi et al. 1994; Helfer & Blitz 1995), and the presence of $H_2O$ masers argues for molecular gas densities $n_{H2} > 10^7\ \mathrm{cm}^{-3}$ (e.g., Kaufman & Neufeld 1996). The shock model applied specifically to component C therefore predicts an emission measure EM ~ $n_e N_e \sim 10^{7\pm1}\ \mathrm{cm}^{-6}$ pc. This shock model prediction is therefore compatible with free-free opacity required to attenuate component C at 1.4 GHz, indicating that the plasma produced by jet-shocked molecular gas could plausibly provide the free-free absorption required to explain the low frequency (< 5 GHz) spectrum of the parsec-scale structure of component C. The proposed geometry all but ensures that the shock region has a clear sight-line to the AGN, and photoionizing radiation may also contribute to the generation of thermal (free-free absorbing) plasma in this region.

## 5.3 Component S2

Measured by MERLIN and the VLA, the high frequency (≥ 5 GHz) spectrum of component S2 is fairly flat: $\alpha(5\text{–}22\ \mathrm{GHz}) = -0.15 \pm 0.08$ on 100-mas scales (Gallimore et al. 1996c). The new VLBA observations are compatible with the flat spectral index: $\alpha(5\text{–}8.4\ \mathrm{GHz}) = -0.20 \pm 0.29$ (Table 3). Broadly consistent with RCWU, the low frequency VLBA spectrum is steep: $\alpha(1.4\text{–}5\ \mathrm{GHz}) = -1.3 \pm 0.1$. Taken with the integrated spectrum, the spectral index map (Figure 6) shows that S2 comprises two components, a compact, ~ 2 mas (0.1 pc) flat-spectrum source surrounded by a larger, ~ 8 mas (0.6 pc) steep-spectrum plasma. This type of structure is just that predicted by the jet-shock scenario we propose for component C (Section 5.2, above), in which case the compact radio source spectrum is flattened locally by free-free absorption through thermal plasma generated by the shock front (Bicknell et al. 1997). The surrounding, steep-spectrum emission may be synchrotron plasma leaking away from the shock front and the region of highest free-free opacity. This model predicts low frequency diminution of the spectrum of the flat-spectrum source, but this prediction will prove difficult to test as the VLBA cannot provide sufficient resolution at low frequency to resolve the compact source from the surrounding steep spectrum emission.

There must be an extended flat-spectrum component on 10-mas scales because ~ 7 mJy of the MERLIN 5 GHz flux density is missing on the tapered VLBA image (Table 3), and the VLA 22 GHz continuum emission is comparably bright, $S_\nu$(22 GHz) ~ 8 mJy (Gallimore et al. 1996c). On the other hand, the source must have a steep spectrum below 5 GHz based on the 1.4 GHz VLBA measurements (Table 3). Again, the simplest solution is a two-component plasma comprising optically-thin synchrotron emission that dominates below 5 GHz and a flat or inverted spectrum source that dominates at 22 GHz. The brightness temperature at 22 GHz is poorly constrained, $5600\ \mathrm{K} < T_b < 5 \times 10^6$ K, limited by the VLA resolution on the low side (Gallimore et al. 1996c) and a VLBA non-detection on the high side (Greenhill et al. 1996). The upper brightness temperature limit rules out synchrotron self-absorption, but these limits allow for the possibility that the 22 GHz flux density may be dominated by thermal free-free emission (Gallimore et al. 1996a). Assuming a temperature of ~ 15,000 K, appropriate for NLR gas, the bolometric luminosity would be < 0.5% of the bolometric luminosity of the AGN. HST imaging reveals copious optical and infrared line emission indicative of $10^4$ K gas in the region of the S1 & S2 sources (Capetti, Axon, & Macchetto 1997a; Thompson et al. 2001), although patchy obscuration may suppress any line emission specifically associated with these compact radio sources. Additional 22 GHz measurements on MERLIN-scale baselines (i.e., intermediate between the VLA and VLBA) should better constrain the brightness temperature as a test of the thermal free-free emission model.

## 5.4 Component NE

The spectrum of component NE is steep both on 100 mas scales (Gallimore et al. 1996c; Muxlow et al. 1996) and on the mas-scale as shown here (Table 3; see also RCWU), and so the emission mechanism is most likely incoherent, optically-thin synchrotron radiation. The VLBA spectral index map is noisy but consistent with uniform spectral index, $\alpha \sim -0.9$, across the source (Figure 6). Based on its morphology and a local spectral flattening observed on VLA and MERLIN images, Gallimore et al. (1996a) proposed that component NE might be associated with a jet shock. RCWU promoted this interpretation because, they argued, the compact VLBA structure is nearly perpendicular to the outflow axis. In fact, component NE orients about 60° from the local jet axis, but the oblique angle does not rule out the shock model since the jet may be impacting material at a glancing angle to the local density gradient. The spectrum of the VLBA radio source is somewhat flatter than the spectrum measured on VLA / MERLIN scales ($\alpha = -1.1$) (Gallimore et al. 1996a), suggesting that reacceleration of the synchrotron particles may be taking place. Scaling from the measured source sizes and flux densities, the equipartition pressure of the VLBA-scale source is ~ 6 times that averaged over the MERLIN-scale emission (as measured at 5 GHz). The implied pressure gradient supports the shock interpretation. There is, however, no evidence indicating that the compact radio components are at or near equipartition except that the implied pressures are in near balance with the surrounding NLR (Gallimore et al. 1996c), and the geometric parameters (magnetic field orientation and optical path length) may change between the VLA / MERLIN scale and the VLBA scale, rendering equipartition ratio calculations very uncertain.

The impact of the jet into a dense NLR cloud should produce a radiative shock in the impacted cloud, resulting in locally enhanced UV continuum and line emission (Taylor, Dyson, & Axon 1992; Dopita & Sutherland 1996; Bicknell et al. 1998). Independent of the shock, photoionizing continuum from the AGN should illuminate the surface of the intruder cloud and produce line emission. On the contrary, there is a lack of strong line emission or optical/UV continuum associated with particularly with component NE (Gallimore et al. 1996a; Capetti et al. 1997b; Kishimoto 1999; Thompson et al. 2001). Selective obscuration cannot easily explain this absence of line emission, because (1) there is no significant enhancement of near-infrared line emission at component NE (Thompson et al. 2001), (2) there is no evidence of attenuation of the UV scattering clouds throughout this region, implying $A_V < 1$ (Kishimoto 1999), and (3) although there is substantial mid-infrared emission associated with NLR clouds, there is no mid-infrared source associated specifically with NE (Bock et al. 2000), meaning any shock related emission has not been reprocessed by intervening, obscuring dust. If NE is a jet shock, the shock cannot have arisen as a result of the jet striking colder, NLR material.

Component NE may instead be nothing more than a local density enhancement in the flow, which may have originated at the start of the jet. There is evidence that the X-ray luminosity of the hidden central engine is variable (Glass 1995; Gallimore et al. 2001; Colbert et al. 2002), suggesting a variable accretion rate (Siemiginowska & Elvis 1997; Janiuk, Czerny, & Siemiginowska 2002). Variable accretion rates presumably might result in variable outflow speeds leading to internal shocks (e.g., Rees 1978; Blandford & Konigl 1979) which will locally enhance the synchrotron emissivity without producing line radiation. The material associated with component NE presumably had to travel through component C, however, where the shock front would compress jet material and magnetic fields causing the pre-shock jet material to lose memory of its longitudinal structure. Alternatively, one could imagine a scenario where a build-up of jet pressure at component C, or perhaps a loss of pressure in the impacted cloud during the evolution of the jet-cloud collision, resulted in a jet breakout that led to the production of component NE.

# 6 Conclusions

The main results of this work are summarized in the following list.

1. On VLBI baselines, the nuclear radio source, S1, resolves into an extended, ~ 0.8 pc long structure oriented nearly perpendicular to the jet axis but more closely aligned to the axis of the $H_2O$ maser disk.
2. On VLBI baselines, component S1 has an overall flat or inverted spectrum between 1.4 and 8 GHz. The emission mechanism cannot be self-absorbed synchrotron radiation because the average brightness temperature is too low. Electron-scattered radiation from a hidden radio nucleus can not contribute significantly to the 5—8.4 GHz continuum without producing significant free-free continuum; the conditions for the scattering model are satisfied only when the free-free flux density is greater than the observed flux density at 8.4 GHz.
3. Component S1 appears to be associated with the inner region of the maser disk, and thermal free-free emission from an X-ray heated disk wind seems to be the most plausible mechanism. To satisfy the heating requirements to produce the observed free-free emission, the hidden AGN luminosity needs to be (at least) $7 \times 10^{44}$ ergs $s^{-1}$, in accord with other independent estimates and models. This model predicts that S1 should be a luminous source of soft X-ray line emission, but this region is obscured from direct view, and the scattered component would contribute insignificantly to the observed soft X-ray spectrum.
4. The compact source within component C appears to be free-free absorbed below 5 GHz (see also RCWU), although unabsorbed, diffuse emission at 1.4 GHz and 5 GHz is detected. The presence of $H_2O$ maser emission suggests that component C arises from a collision between the radio jet and a molecular cloud (Gallimore et al. 1996a; Gallimore et al. 1996b), and we propose that the free-free absorption might arise from an ionized cocoon produced by the shock (cf. Bicknell et al. 1997).
5. Component C also appears to be variable at 5 GHz. We speculate that the variability may owe to the evolution of the jet shock.
6. Component S2 is peculiar in that it has a steep spectrum between 1.4 and 5 GHz but a relatively flat spectrum between 5 and 8.4 GHz. The peculiar spectrum may result from spatial filtering of an extended, steep-spectrum component at higher frequencies. The compact flat-spectrum component may mark a free-free absorbed shock region of the jet akin to Component C.
7. Component NE, which remains steep-spectrum on VLBA baselines, appears to be a local emissivity enhancement in the radio jet. No optical emission-line clouds are found associated specifically with this component, and so there is no independent evidence for the presence of a shock between the jet and the surrounding ISM. Component NE may mark the location of an internal shock, or perhaps a distinct blob of jet plasma that broke away from an earlier shock (component C?) in the jet.

**Acknowledgements**

JFG was supported by a Jansky fellowship at NRAO-Charlottesville during the formative work on this project. JFG received further support from a DDRF grant for research leave at STScI. Ali Kinkhabwala kindly provided an electronic version of the XMM-Newton RGB spectrum. This work greatly benefited from conversations with Moshe Elitzur, Hagai Netzer, Martin Elvis, & Julian Krolik. We also thank an anonymous referee who pointed out some technical errors and made helpful suggestions that improved the error analysis.

## Figures

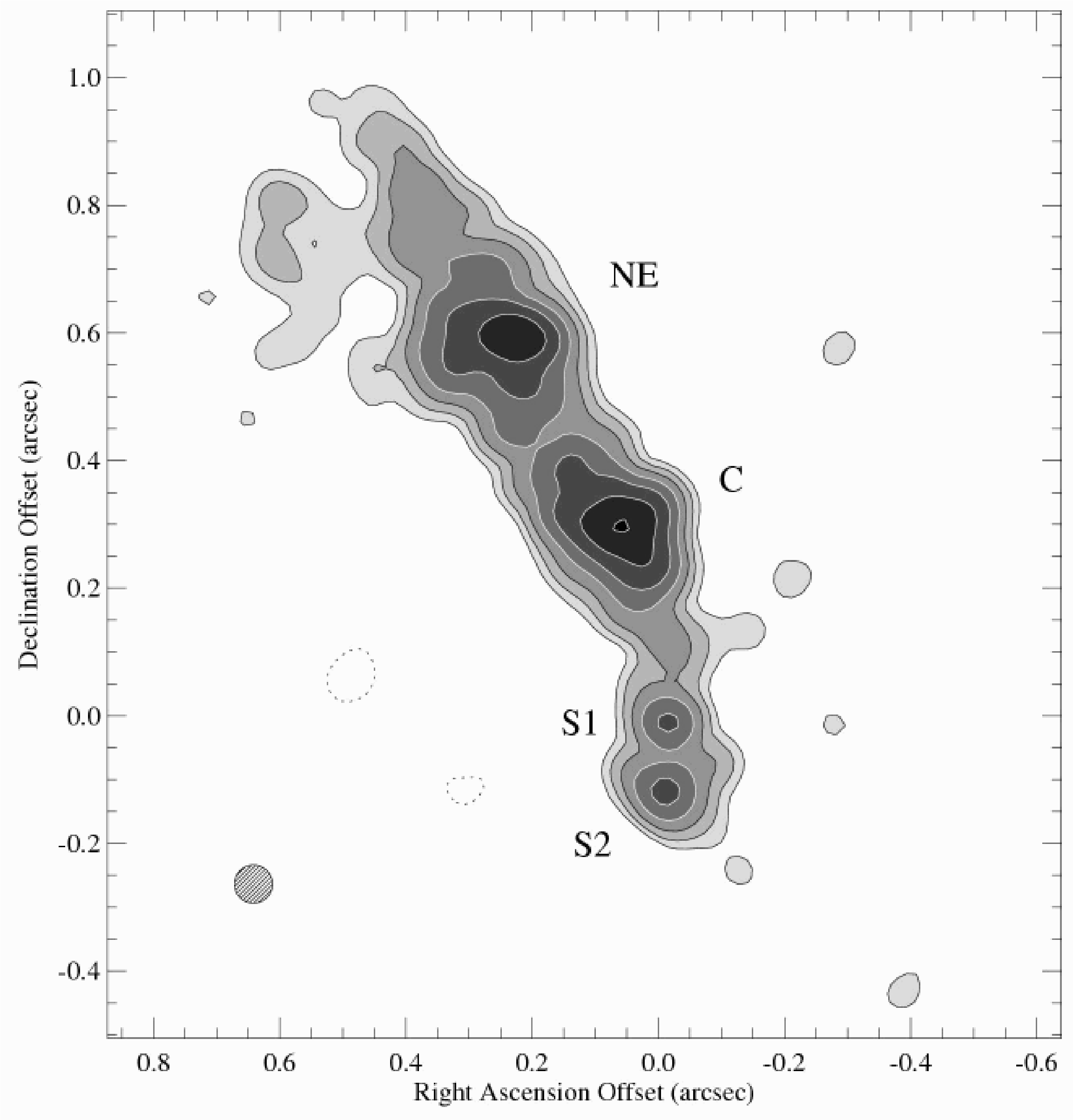

Figure 1. New MERLIN 5 GHz contour image of NGC 1068. The axis coordinates are measured with respect to the centroid position of component S1 on the new VLBA 5 GHz continuum image: RA(2000) = $2^h\ 42^m\ 40.70905^s$, Dec(2000) = –0° 0′ 47.945″. The logarithmically-spaced contour levels are: ±0.46, 0.98, 2.1, 4.5, 9.7, 20.8, & 44.6 mJy beam$^{-1}$. The annotations give the names of the brightest compact radio sources. The FWHM size of the CLEAN restoring beam is indicated by the cross-hatched ellipse in the lower left corner of the figure. For the distance assumed in this paper, 1″ = 72 pc.

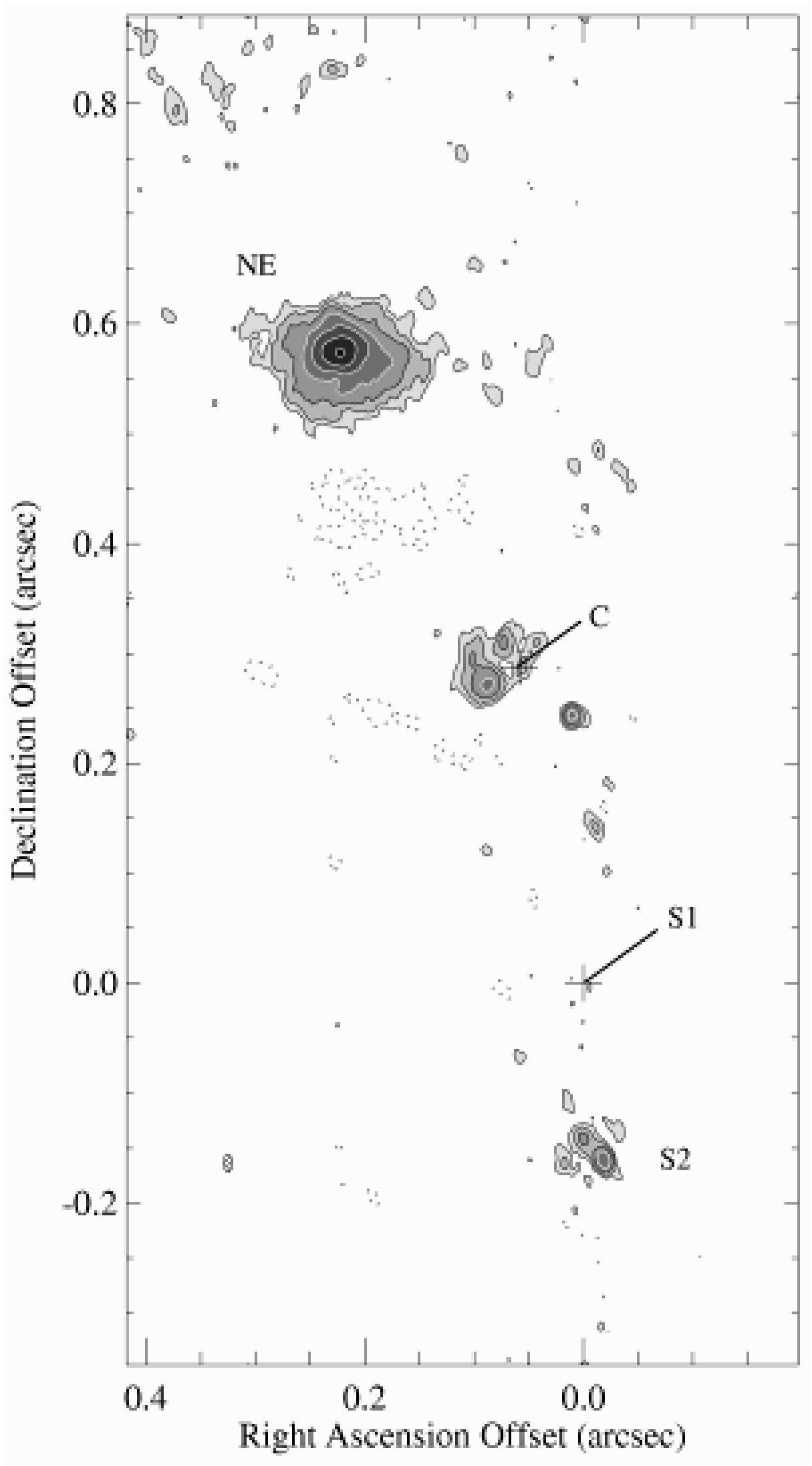

Figure 2. VLBA + phased-VLA 1.4 GHz contour image of NGC 1068. The plotting convention is identical to Figure 1. The logarithmically-spaced contour levels are: ±0.8, 1.3, 2.0, 3.2, 5.0, 7.9, & 12.5 mJy beam$^{-1}$. The VLBA 5 GHz positions of S1 and C are marked by "+" symbols and annotation.

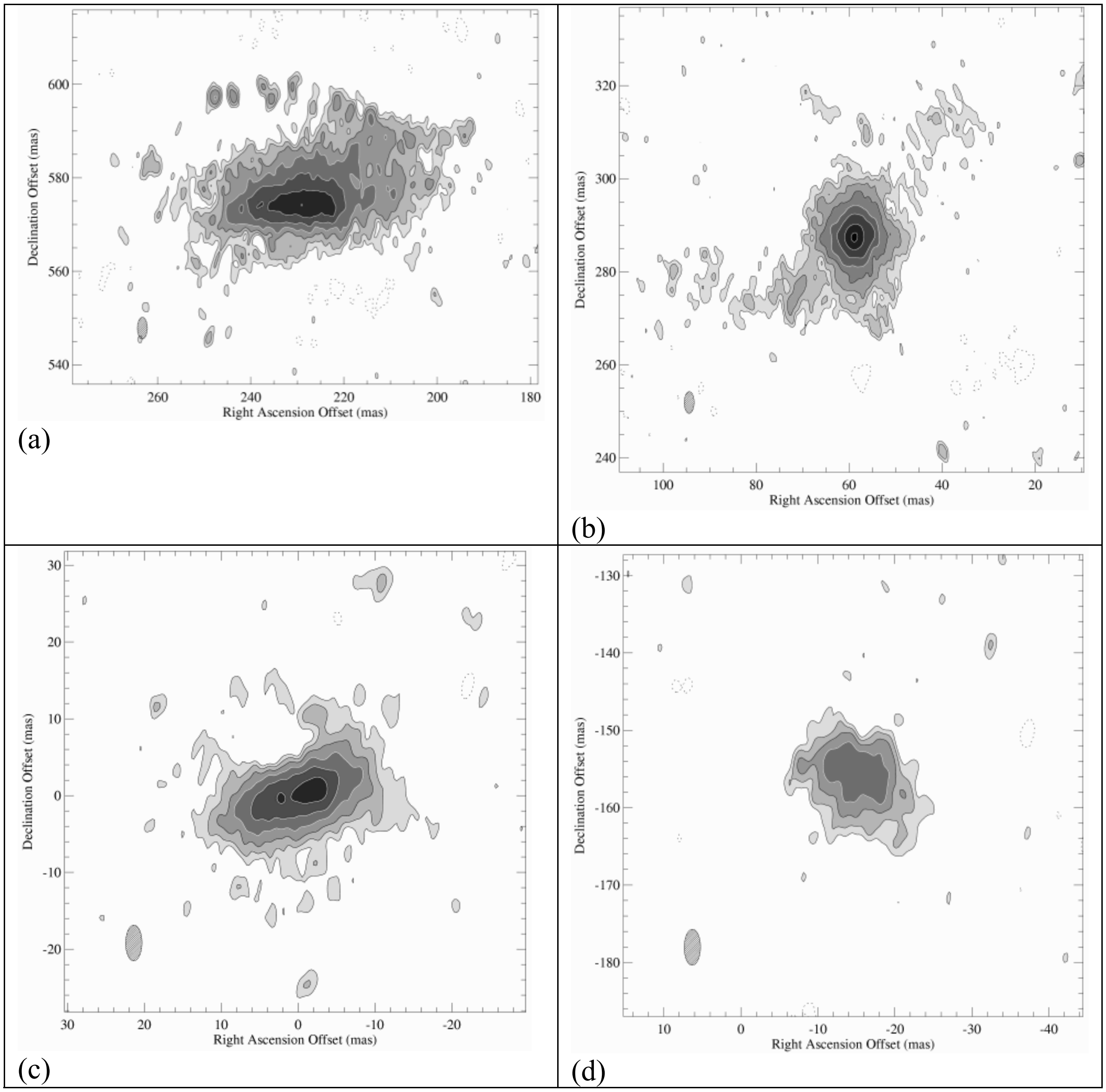

Figure 3. VLBA 5 GHz contour images of NGC 1068. (a) Component NE. (b) Component C. (c) Component S1. (d) Component S2. The coordinate convention is the same as in Figure 1. The logarithmically-spaced contour levels are: ±0.11, 0.16, 0.24, 0.35, 0.51, 0.75, 1.10, & 1.50 mJy $beam^{-1}$ (the highest contour appears only on the image of component C). For the distance assumed in this paper, 1 pc = 14 mas.

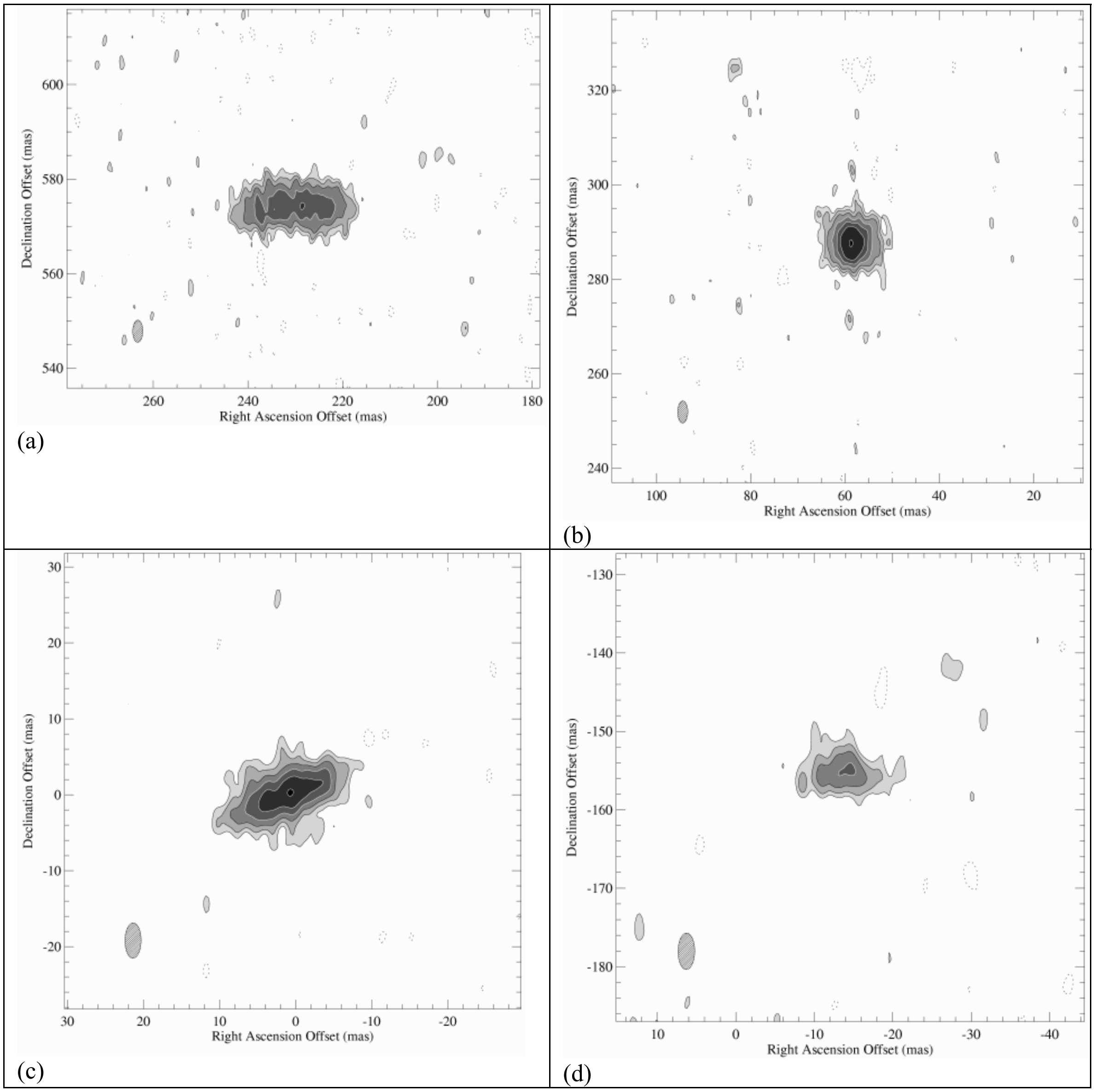

Figure 4. VLBA 8.4 GHz contour images of NGC 1068. (a) Component NE. (b) Component C. (c) Component S1. (d) Component S2. The coordinate convention is the same as in Figure 1. The logarithmically-spaced contour levels are: ±0.13, 0,20, 0.29, 0.44, 0.66, 1.0, & 2.44 mJy beam$^{-1}$ (the highest contour appears only on the image of component C).

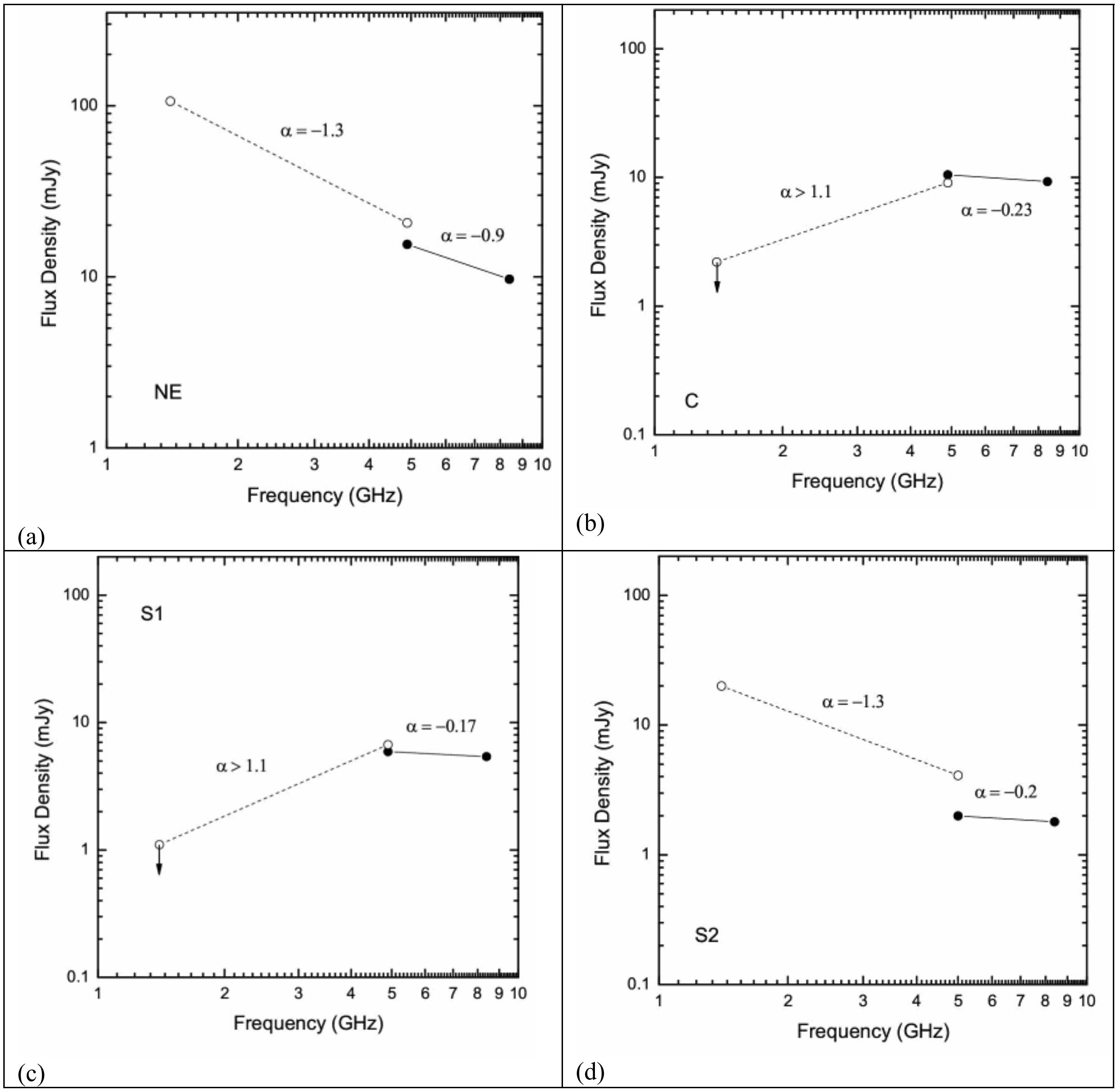

Figure 5. VLBA radio continuum spectra of the compact radio sources of NGC 1068. (a) Component NE. (b) Component C. (c) Component S1. (d) Component S2. Open symbols and dotted lines mark (*u*,*v*)-filtered and masked measurements at 1.4 and 5 GHz at the resolution of the 1.4 GHz images (16 × 8 mas), and closed symbols connected by solid lines mark the masked measurements at 5 GHz and 8.4 GHz at the resolution of the 5 GHz images (5 × 2 mas) (see Section 3.1 for a more detailed description). The upper limits for non-detections (Components S1 and C at 1.4 GHz) are $90^{th}$ percentile.

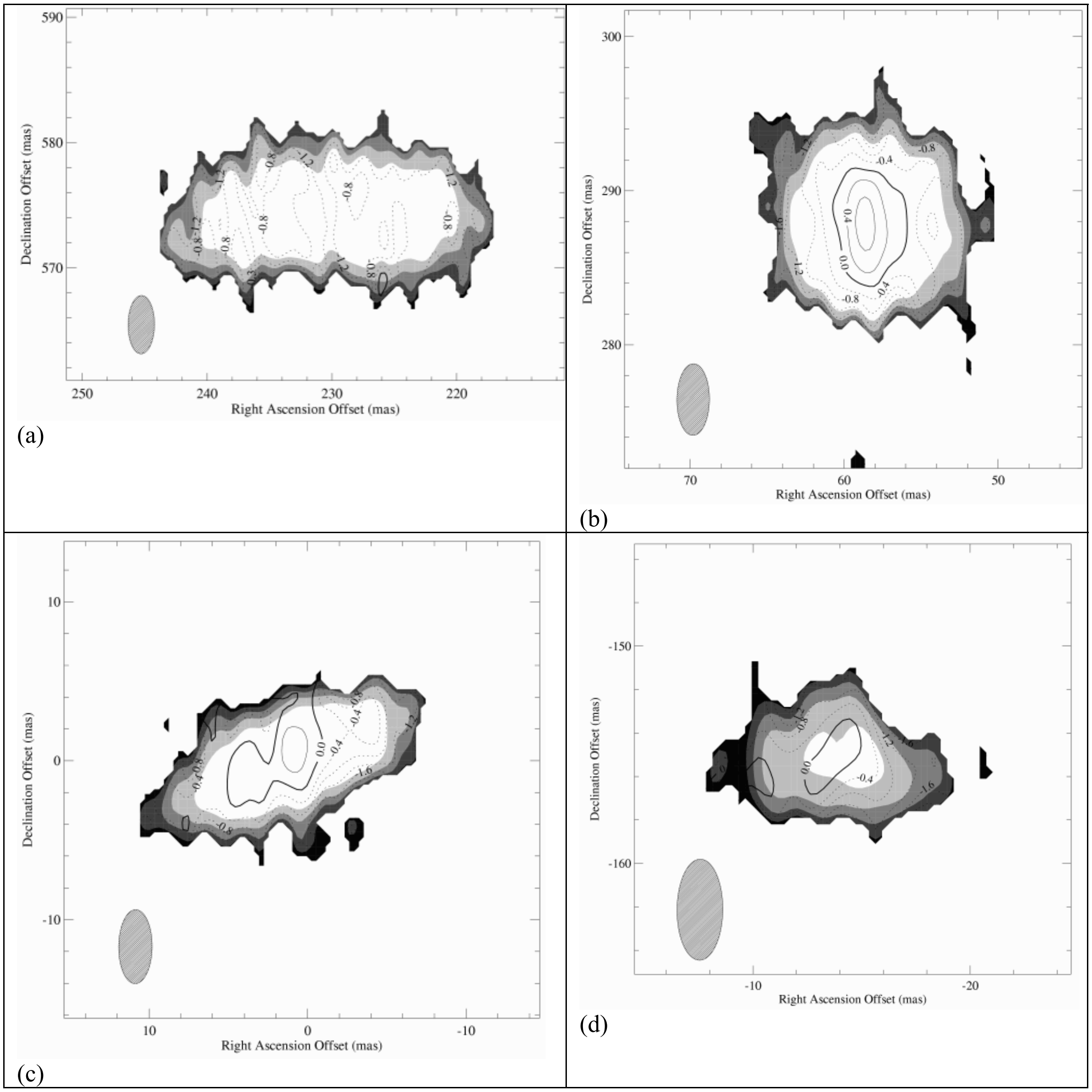

Figure 6 The 5−8.4 spectral index distribution of the compact radio sources of NGC 1068. (a) component NE. (b) component C. (c) component S1. (d) component S2. The spectral indices are represented by annotated contours, and the uncertainties are indicated by the shaded regions. The contour levels for (a) are $\alpha = -1.2, -0.8, -0.3$, & 0, and the contour levels for (b−d) are $\alpha = -1.2$ to $+1.2$ in steps of 0.4. Negative (steep) spectral indices are indicated by dotted contours, and positive (inverted) contours are indicated by solid contours, with $\alpha = 0$ marked by the heavy contour. The uncertainty shades step from $\Delta\alpha < 0.2$ (white), 0.2−0.4, 0.4−0.8, and > 0.8 (dark). Note that the uncertainty of the flux scale and the error introduced by deconvolution introduce an overall uncertainty of order 0.1 – 0.2 dex.

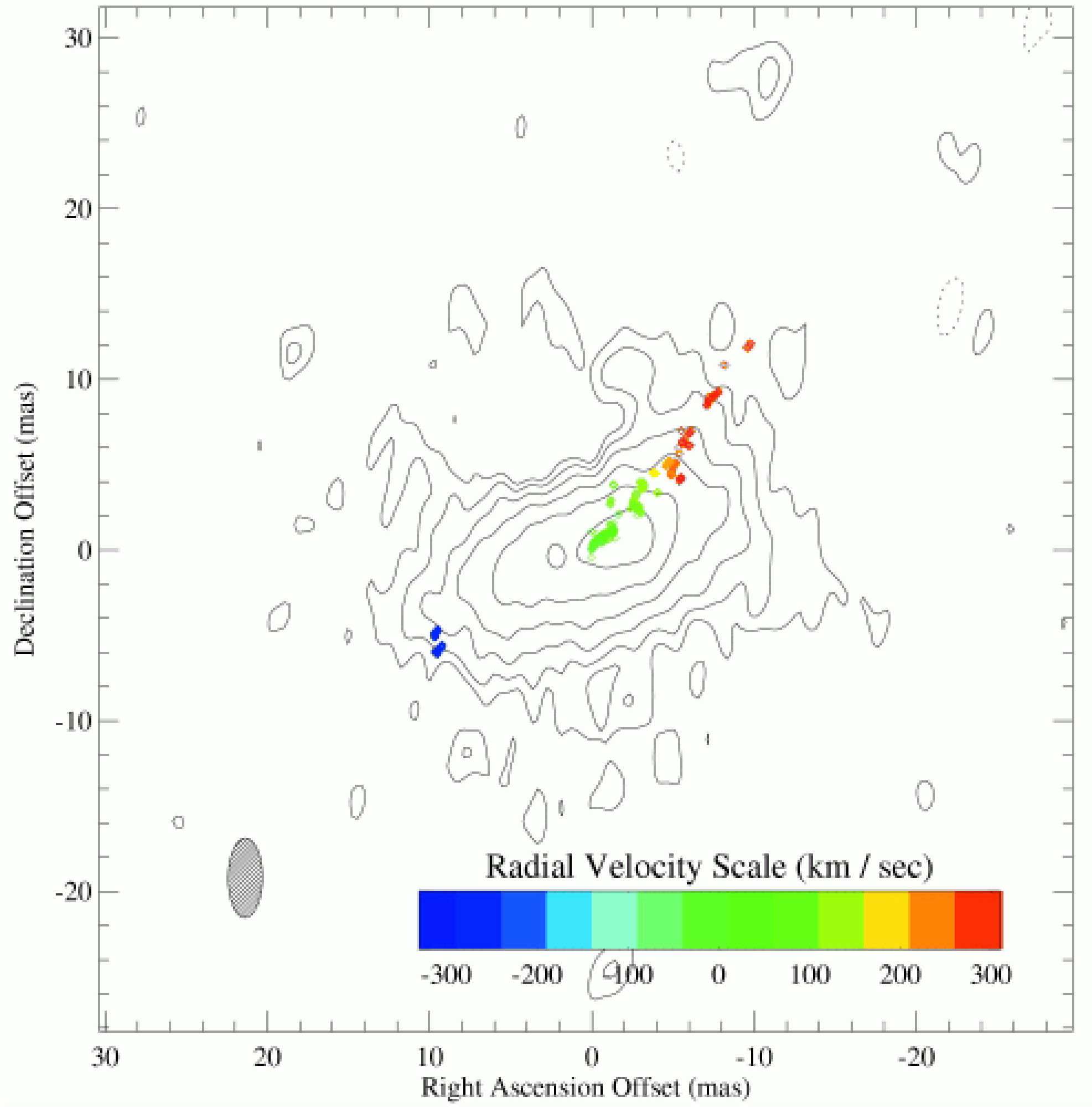

Figure 7. The location of the S1 maser spots relative to the VLBA 5 GHz radio continuum images in contours, with contour levels as in Figure 3. The maser spot positions are plotted as compact, open diamonds, which are color-coded by the radial velocity. The color scale bars at the bottom of the figure gives the radial velocities relative to the systemic velocity of the host galaxy, 1150 km $s^{-1}$. The uncertainty of the alignment is 1.3 mas along the long axis of the maser distribution and 1.1 mas perpendicular to the long axis.

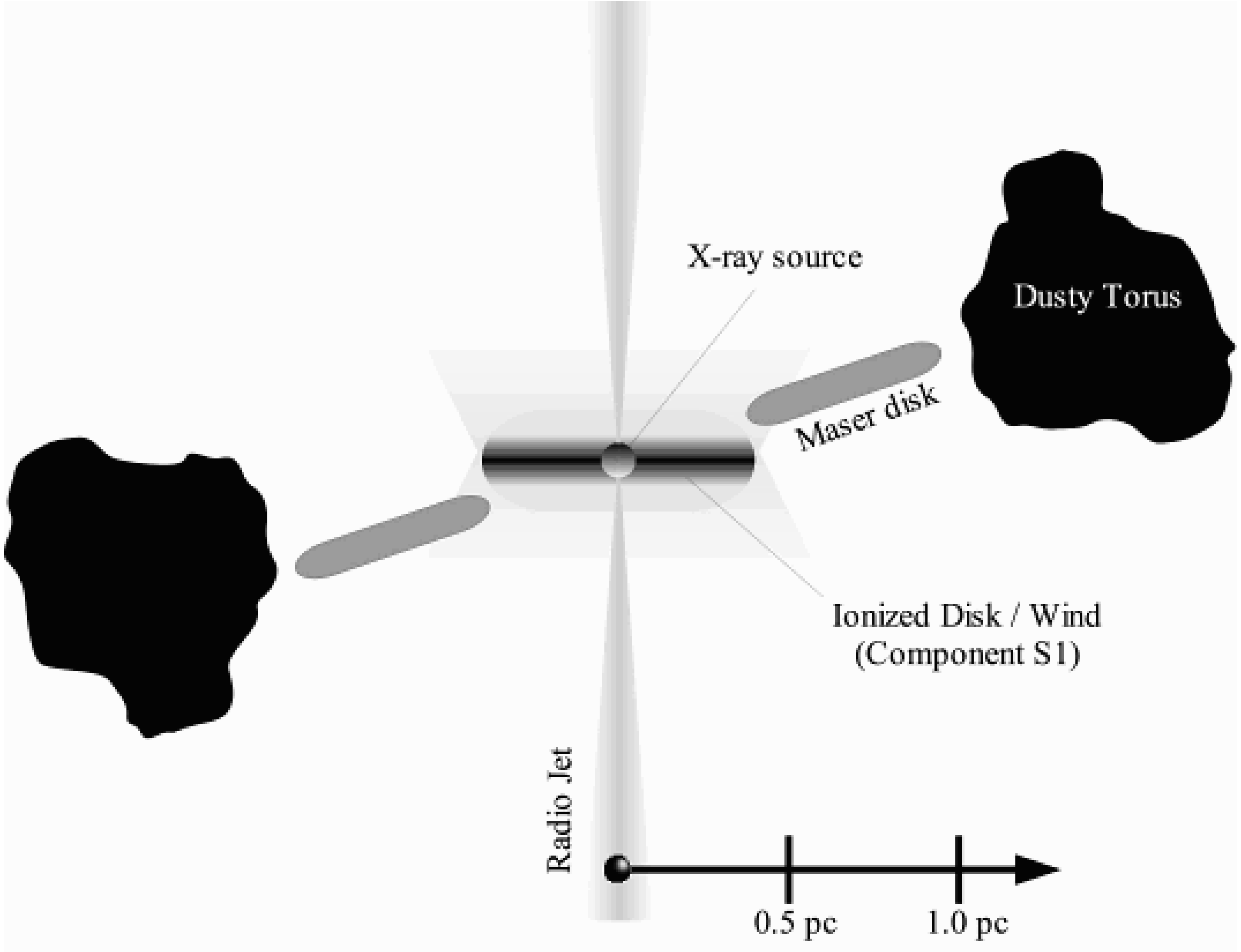

Figure 8: A schematic diagram for the disk model of component S1. The central parsec is shown in a cutaway view, with the S1 disk occupying the inner 0.4 pc (based on the present work), the maser disk spanning 0.6 – 1 pc (Greenhill & Gwinn 1997), and the inner surface of the dusty, obscuring torus located at ~ 1 pc (Thompson & Corbin 1999; Jaffe et al. 2004). There is a slight misalignment between the S1 radio axis and the line traced by the $H_2O$ maser spots. The geometry of the dusty torus is essentially unknown, except that clouds in the torus must block our direct view of the central parsec.

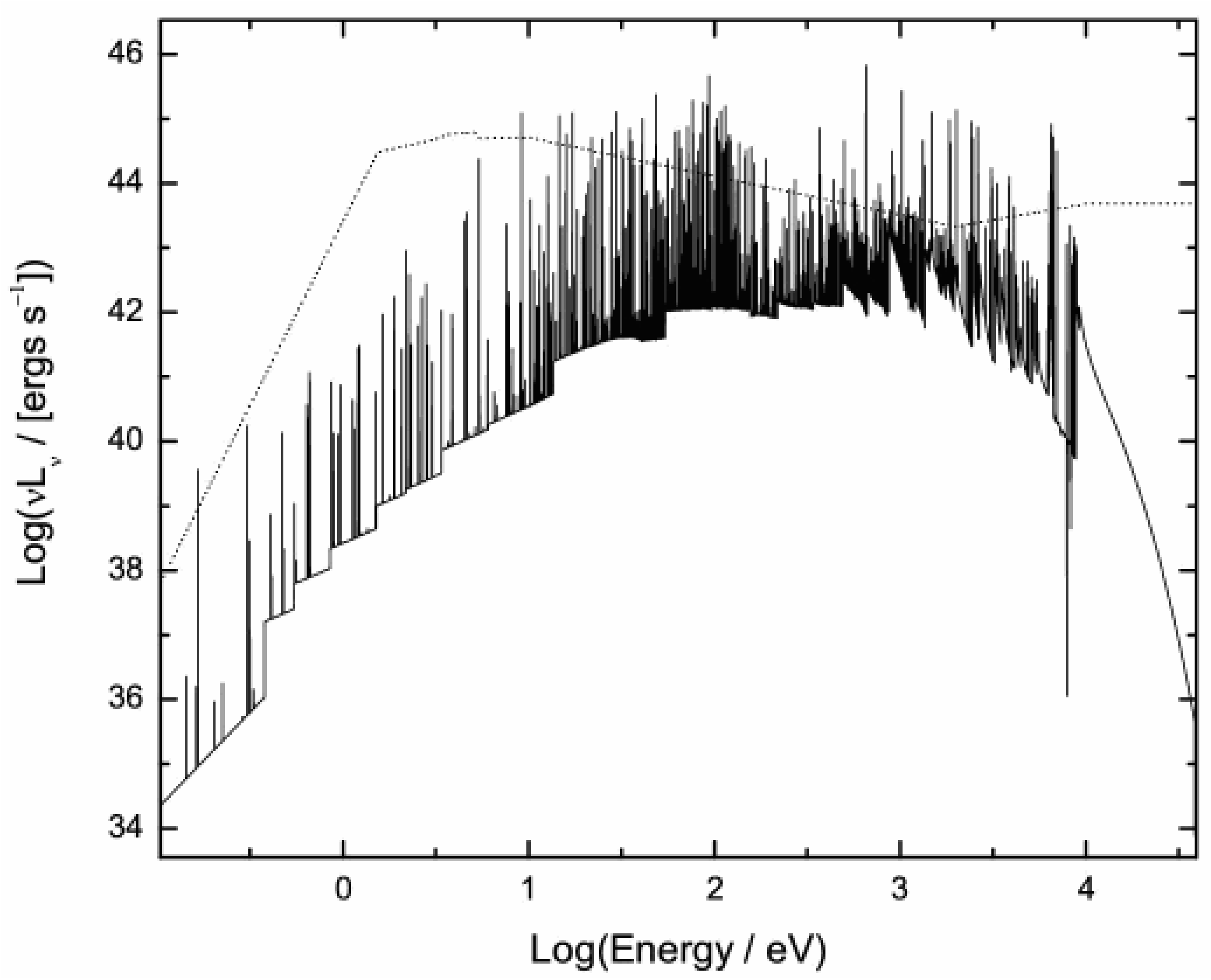

Figure 9. The optical—X-ray AGN spectral energy distribution (dotted line) and the predicted spectrum of S1 (solid line) based on XSTAR models. The AGN SED is normalized to $L_{AGN}$(0.1 eV – 100 keV) = 7 × $10^{44}$ ergs $s^{-1}$, i.e., the minimum luminosity required to produce the observed 8.4 GHz flux density of S1 (see 5.1.3 for details), and the S1 spectrum is the result of the minimum luminosity model.

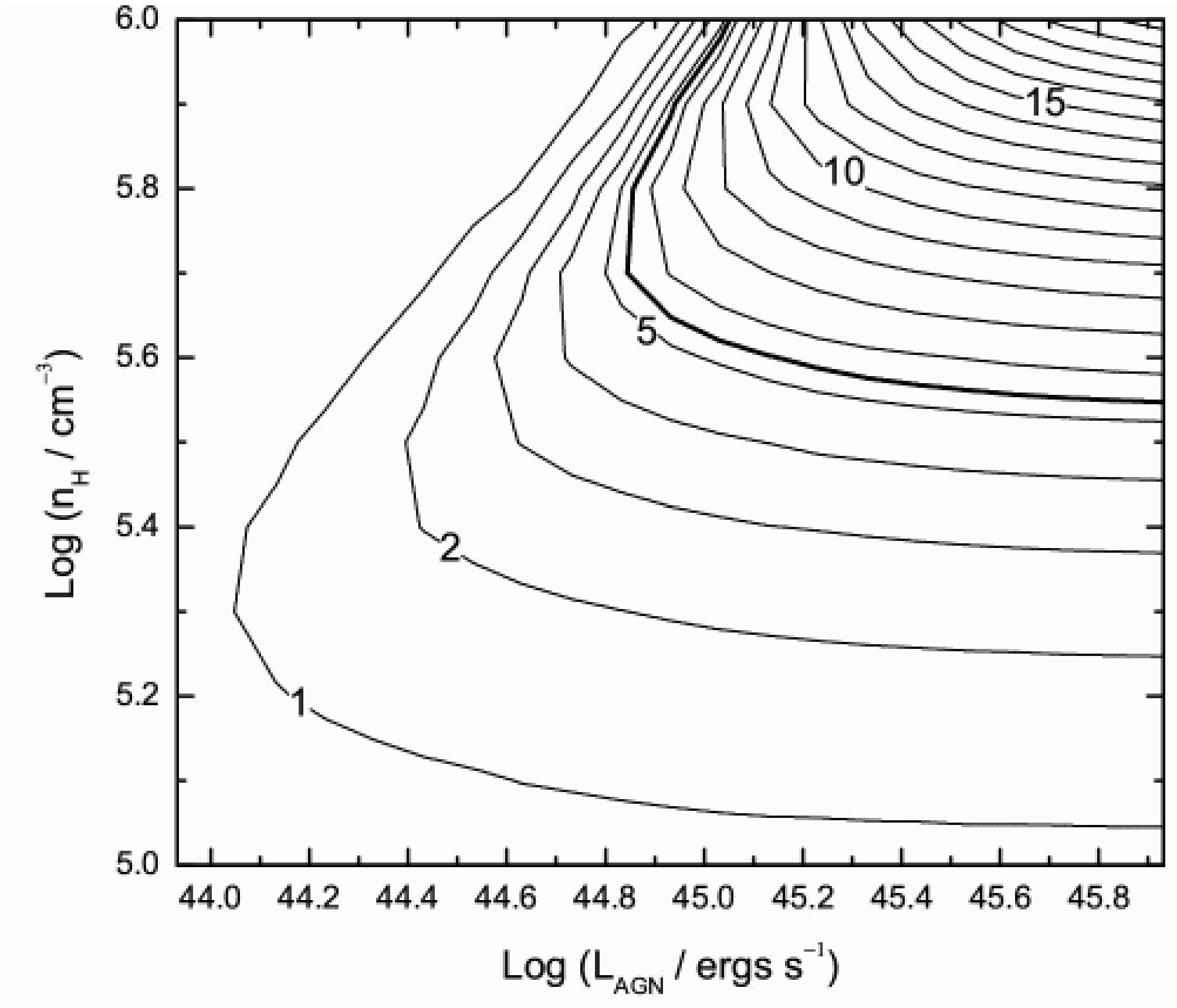

Figure 10 Results of X-ray heating models for component S1. The contours trace the predicted thermal free-free 8.4 GHz emission, in mJy, from component S1 as a function of plasma density and AGN luminosity. The thin contour increment is 1 mJy. The measured 8.4 GHz flux is 5.4 mJy, marked by the thicker contour line.

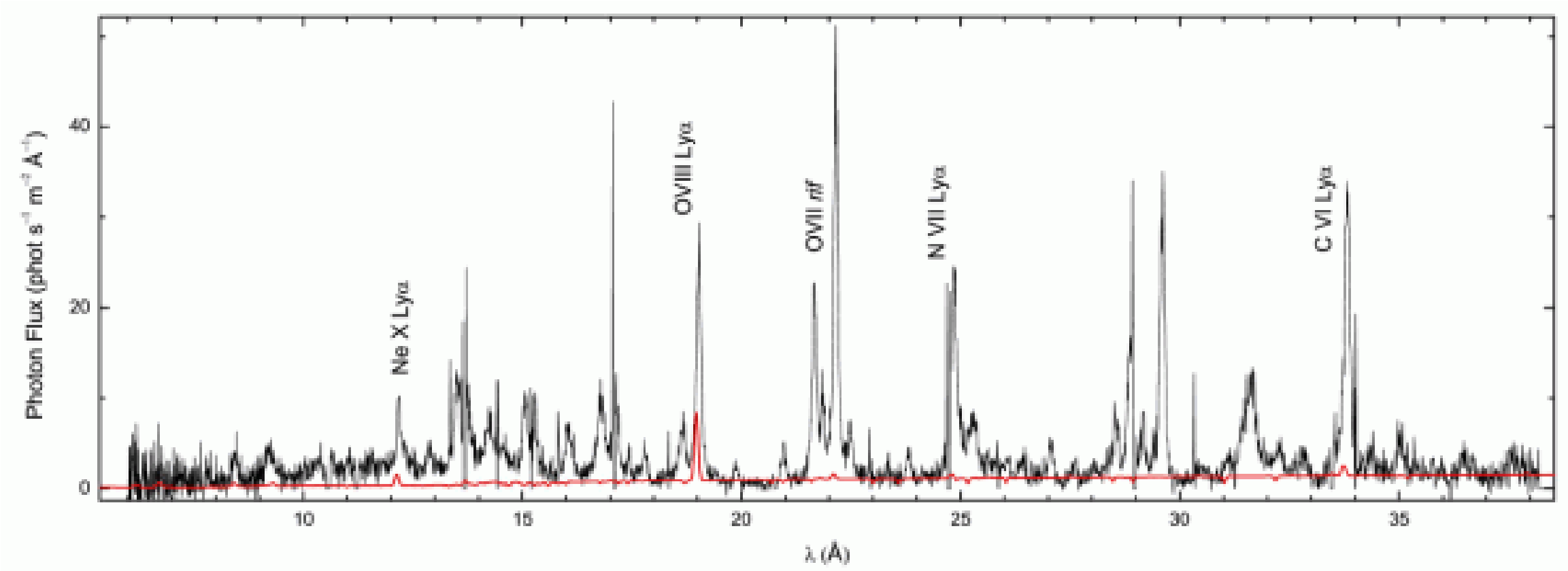

Figure 11 Comparison of the XMM-Newton RGS spectrum of NGC 1068 (dark line) with the soft X-ray model spectrum for the AGN (based on Pier et al. 1994) and S1 disk region (red line). The XSTAR model was scaled down to match the average continuum level of the XMM-Newton spectrum. The scaling factor, 0.12%, sets an upper limit on the contribution of electron-scattered radiation from the AGN to the soft X-ray band. The brightest model lines, commonly Lyα transitions of hydrogenic metals, are labeled.

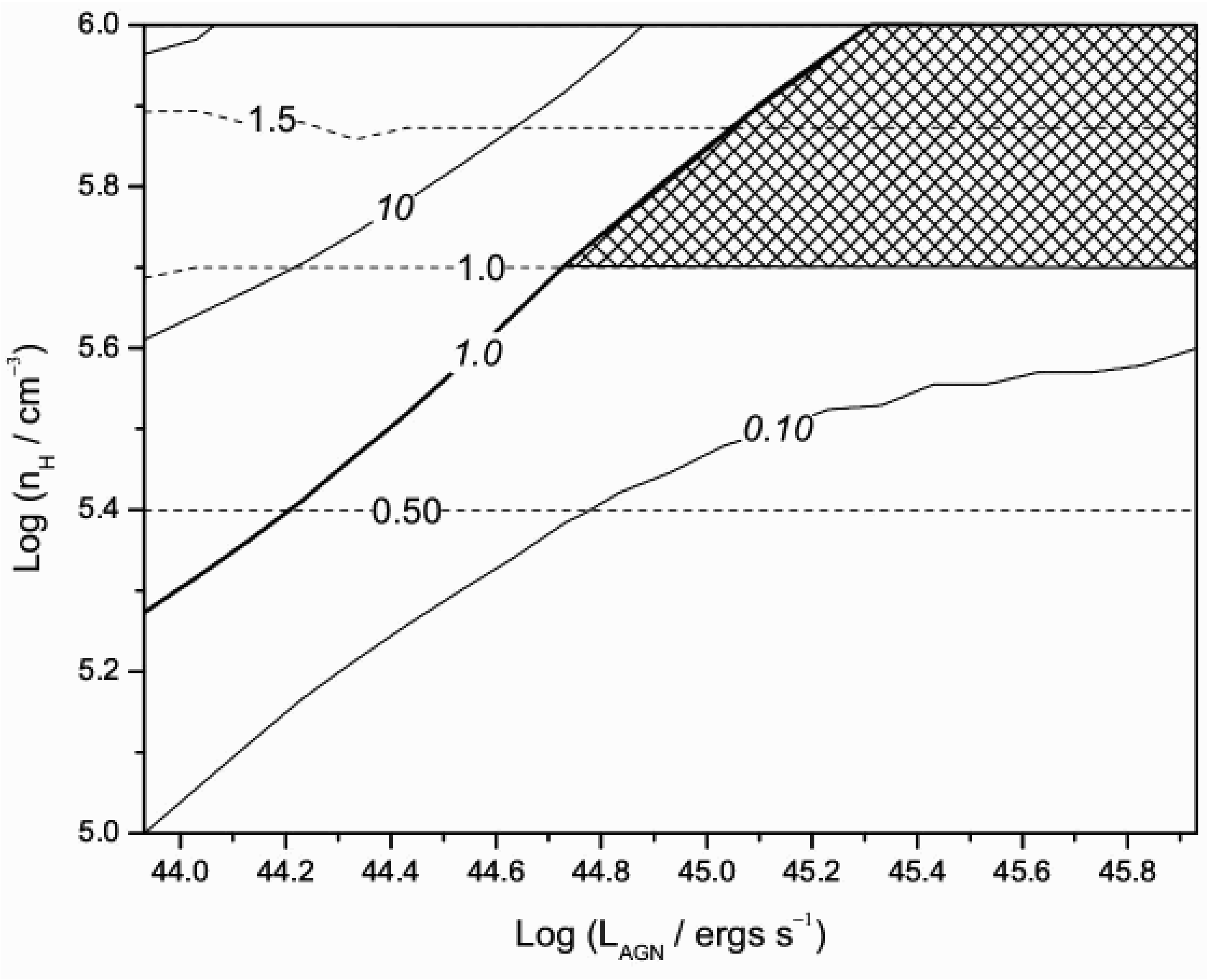

Figure 12 Grids of 8.4 GHz free-free opacities (solid lines with italic labels) and electron scattering opacities (dashed lines with regular labels) as functions of AGN luminosity and plasma density derived from the X-ray heating models for S1. The crosshatched area indicates the region where the scattering opacity exceeds unity and the free-free opacity is less than unity, as would be required for a scattered compact radio source to dominate the 8.4 GHz flux density.

# Tables

**Table 1 Image Properties**

| Telescope | Band (GHz) | Integration Time (hrs) | RMS (μJy beam$^{-1}$) | RMS ($10^4$ K) | Beam Size (mas × mas, deg) | | |
|---|---|---|---|---|---|---|---|
| MERLIN | 5.0 (C) | 7.61 | 190 | 0.17 | 103.1 | 51.6 | 20.5 |
| VLBA | 1.4 (L) | 5.31 | 21 | 11 | 16.0 | 7.6 | 1.0 |
| VLBA | 5.0 (C) | 5.26 | 37 | 16 | 4.7 | 2.1 | –0.5 |
| VLBA | 8.4 (X) | 8.15 | 50 | 8 | 4.7 | 2.1 | –0.5 |

**Table 2. Image Moment Analysis of the VLBA Radio Components**

| Component | ν (GHz) | $S_\nu$ (mJy) | | Major Axis (mas) | | Minor Axis (mas) | | PA (°) | | RA (s) | Dec (″) | Err (mas) |
|---|---|---|---|---|---|---|---|---|---|---|---|---|
| NE | 1.4 | 283.2 | (22.8) | 73.2 | (13.0) | 54.4 | (8.4) | 84.0 | (0.5) | 40.72376 | –47.3728 | 2.0 |
| NE | 5.0 | 36.7 | (4.1) | 30.8 | (1.8) | 18.8 | (1.3) | 103.5 | (2.2) | 40.72426 | –47.3685 | 0.8 |
| NE | 8.4 | 9.7 | (0.8) | 19.6 | (2.3) | 7.1 | (0.4) | 90.6 | (0.7) | 40.72440 | -47.3708 | 0.1 |
| C | 1.4 | 42.3 | (3.7) | 62.4 | (4.9) | 43.1 | (3.0) | –42.0 | (8.9) | 40.71476 | –47.6599 | 3.5 |
| C | 5.0 | 19.4 | (1.6) | 16.7 | (3.9) | 13.2 | (1.9) | –37.6 | (16.6) | 40.71298 | –47.6577 | 0.5 |
| C | 8.4 | 9.3 | (0.8) | 6.1 | (1.0) | 5.7 | (1.1) | – | – | 40.71295 | –47.6573 | 0.1 |
| S1 | 1.4 | <0.06 | – | – | – | – | – | – | – | – | – | – |
| S1 | 5.0 | 9.1 | (0.8) | 16.6 | (0.2) | 11.2 | (2.4) | 104.5 | (3.9) | 40.70907 | –47.9445 | 0.4 |
| S1 | 8.4 | 5.4 | (0.5) | 11.0 | (0.4) | 3.7 | (0.8) | 108.1 | (1.4) | 40.70912 | –47.9449 | 0.1 |
| S2 | 1.4 | 20.0 | (1.9) | 41.7 | (1.8) | 27.1 | (3.7) | 58.2 | (6.2) | 40.70823 | –48.1003 | 2.9 |
| S2 | 5.0 | 4.1 | (0.4) | 11.8 | (0.2) | 7.0 | (0.6) | 56.1 | (1.0) | 40.70803 | –48.1010 | 0.2 |
| S2 | 8.4 | 1.8 | (0.2) | 8.8 | (0.1) | 3.1 | (0.3) | 85.3 | (2.4) | 40.70813 | –48.1001 | 0.1 |

**Table 3 Spectral Index Measurements**

| Component | $S_{1.4}$, masked (mJy) | | $S_5$, tapered (mJy) | | α(1.4–5) | | $S_5$, masked (mJy) | | α(5–8.4) | |
|---|---|---|---|---|---|---|---|---|---|---|
| NE | 106.5 | (8.5) | 20.7 | (1.7) | –1.3 | (0.1) | 15.5 | (1.3) | –0.90 | (0.22) |
| C | <2.2 | – | 9.1 | (0.7) | > 0.9 | – | 10.5 | (0.9) | –0.23 | (0.23) |
| S1 | <0.7 | – | 5.9 | (0.5) | > 1.5 | – | 5.9 | (0.5) | –0.17 | (0.24) |
| S2 | 16 | (1.6) | 3.0 | (0.3) | –1.3 | (0.1) | 2.0 | (0.2) | –0.20 | (0.29) |